\documentclass[preprint2]{aastex6}

\usepackage{times}

\newcommand{\Msun}{\ensuremath{\,{\rm M}_\odot}}                  % Solar mass symbol
\newcommand{\Rsun}{\ensuremath{\,{\rm R}_\odot}}                  % Solar radius symbol
\newcommand{\Teff}{\ensuremath{T_{\rm eff}}}                      % Effective temperature symbol
\newcommand{\Mearth}{\ensuremath{\,{\rm M}_\oplus}}               % Earth mass symbol
\newcommand{\Rearth}{\ensuremath{\,{\rm R}_\oplus}}               % Earth radius symbol
\newcommand{\Teq}{\ensuremath{T_{\rm eq}^{\,\prime}}}             % Equilibrium temperature symbol
\newcommand{\ms}{\,m\,s$^{-1}$}                                   % m/s^2 symbol
\newcommand{\mss}{\,m\,s$^{-2}$}                                  % m/s^2 symbol
\newcommand{\as}{\ensuremath{^{\prime\prime}}}                    % Arcsecond symbol
\newcommand{\am}{\ensuremath{^\prime}}                            % Arcminute symbol
\newcommand{\FeH}{\ensuremath{\left[\frac{\rm Fe}{\rm H}\right]}} % [Fe/H] symbol
\newcommand{\psun}{\ensuremath{\,\rho_\odot}}                     % Solar density symbol
\newcommand{\chir}{\ensuremath{\chi_\nu^{\,2}}}                   % Reduced chi-squared symbol
\newcommand{\mc}[1]{\multicolumn{2}{c}{#1}}

\newcommand{\er}[3]{\ensuremath{#1^{+#2}_{-#3}}}

\newcommand{\reff}[1]{{#1}}

\begin{document} %%%%%%%%%%%%%%%%%%%%%%%%%%%%%%%%%%%%%%%%%%%%%%%%%%%%%%%%%%%%%%%%%%%%%%%%%%%%%%%%%%%%%%%%%%%%%%%%%%%%%%%%%%%%%%%%%%%%%%%%%%%%%%%%%%%%
%%%%%%%%%%%%%%%%%%%%%%%%%%%%%%%%%%%%%%%%%%%%%%%%%%%%%%%%%%%%%%%%%%%%%%%%%%%%%%%%%%%%%%%%%%%%%%%%%%%%%%%%%%%%%%%%%%%%%%%%%%%%%%%%%%%%%%%%%%%%%%%%%%%%%

\title{Detection of the atmosphere of the 1.6 Earth mass exoplanet GJ\,1132\,b}

\author{John Southworth\,$^{1}$, Luigi Mancini\,$^{2,3,4}$, Nikku Madhusudhan$^{5}$, Paul Molli\`ere\,$^{2}$, Simona Ciceri$^{6}$, Thomas Henning\,$^{2}$}

\affil{$^{1}$\,Astrophysics Group, Keele University, Staffordshire, ST5 5BG, UK \\
        $^{2}$\,Max Planck Institute for Astronomy, K\"onigstuhl 17, 69117 Heidelberg, Germany \\
        $^{3}$\,Dipartimento di Fisica, Universit\`a di Roma Tor Vergata, Via della Ricerca Scientifica 1, 00133 -- Roma, Italy \\
        $^{4}$\,INAF -- Osservatorio Astrofisico di Torino, via Osservatorio 20, 10025, Pino Torinese, Italy \\
        $^{5}$\,Institute of Astronomy, University of Cambridge, Madingley Road, Cambridge CB3 0HA, UK \\
        $^{6}$\,Department of Astronomy, Stockholm University, SE-106 91 Stockholm, Sweden}

\begin{abstract}
Detecting the atmospheres of low-mass low-temperature exoplanets is a high-priority goal on the path to ultimately detect biosignatures in the atmospheres of habitable exoplanets. High-precision HST observations of several super-Earths with equilibrium temperatures below 1000\,K have to date all resulted in featureless transmission spectra, which have been suggested to be due to high-altitude clouds. We report the detection of an atmospheric feature in the atmosphere of a 1.6\Mearth\ transiting exoplanet, GJ\,1132\,b, with an equilibrium temperature of $\sim$600\,K and orbiting a nearby M dwarf. We present observations of nine transits of the planet obtained simultaneously in the $griz$ and $JHK$ passbands. We find an average radius of $1.43\pm0.16$\Rearth\ for the planet, averaged over all the passbands, \reff{and a radius of $0.255\pm0.023$\Rsun\ for the star, both of which are significantly greater than previously found. The planet radius} can be decomposed into a ``surface radius'' at $\sim$1.375\Rearth\ \reff{overlaid by atmospheric features which increase the observed radius in the $z$ and $K$ bands}. The $z$-band radius is 4$\sigma$ higher than the continuum, suggesting a strong detection of an atmosphere. We deploy a suite of tests to verify the reliability of the transmission spectrum, which are greatly helped by the existence of repeat observations. The large $z$-band transit depth indicates strong opacity from H$_2$O and/or CH$_4$ or a hitherto unconsidered opacity. A surface radius of $1.375\pm0.16$\Rearth\ allows for a wide range of interior compositions ranging from a nearly Earth-like rocky interior, with $\sim$70\% silicate and $\sim$30\% Fe, to a substantially H$_2$O-rich water world. %New observations with HST and existing ground-based facilities would be able to confirm the present detection and further constrain the atmospheric composition of the planet.
\end{abstract}


%%%%%%%%%%%%%%%%%%%%%%%%%%%%%%%%%%%%%%%%%%%%%%%%%%%%%%%%%%%%%%%%%%%%%%%%%%%%%%%%%%%%%%%%%%%%%%%%%%%%%%%%%%%%%%%%%%%%%%%%%%%%%%%%%%%%%%%%%%%%%%%%%%%%%

\section{Introduction}                                                                                                              \label{sec:intro}

M dwarfs are bounteous throughout our Galaxy, and host large numbers of planets \citep{Cassan+12nat,DressingCharbonneau15apj}. The myriad of planets orbiting M dwarfs is dominated by small and low-mass rocky bodies with masses and radii comparable to Earth's: \citep{MortonSwift14apj,Gaidos+16mn}, and many occur in multiple systems \citep{Muirhead+15apj}. Planets around M dwarfs are of particular interest because the dimness of the host stars means their habitable zones \citep[e.g.][]{Kopparapu+13apj} are located at short orbital periods, which are much more accessible to observational study \citep{Gillon+16nat}. \reff{However, planets within the habitable zone of M dwarfs face additional challenges such as tidal locking and large incident fluxes of high-energy photons \citep{Lammer+03apj,Cunha++14ijasb,Shields+16}.}

From an analysis of the M dwarfs observed by the {\it Kepler} satellite, \citet{Gaidos+16mn} found that there were on average 2.2 planets per star and that the radius distribution peaked at 1.2\Rearth. These objects are expected to have tenuous atmospheres, with those of radius below approximately 1.5--1.6\Rearth\ being almost entirely rocky \citep{WeissMarcy14apjl,Rogers15apj}. This raises a problem for the overarching goal of constraining the chemical compositions and atmospheric characteristics of rocky planets, because the observational signatures of their atmospheres are below the levels detectable with current \reff{facilities}.

GJ\,1132 is a nearby very-low-mass star which was recently found to host a transiting and potentially rocky planet \citep[][hereafter BT15]{Berta+15nat} of mass 1.6\Mearth\ and radius 1.2\Rearth. Its proximity to the Sun ($12.04\pm0.24$\,pc; \citealt{Jao+05aj}) means that it is comparatively bright and therefore well-suited to analyses aimed at constraining the properties of both the star and the planet. \citet{Schaefer+16apj} presented simulations of the interior and atmosphere of GJ\,1132\,b, finding that the atmosphere is likely tenuous and dominated by O$_2$. Whilst significantly hotter than Earth, the planet is one of the coolest transiting planets of known mass and is therefore of great interest for comparative planetology. BT15 found that GJ\,1132\,A is an old (5\,Gyr or more) and slowly-rotating (rotation period 125\,d) M4.5\,V star, making it representative of a large population of planet host stars expected to be found in the (relatively) near future \citep{Cloutier+17aj}.

\reff{\citet[][hereafter D16]{Dittmann+16} presented extensive photometry of the GJ\,1132 system, comprising 21 transits observed with the MEarth telescopes and a 100\,hr light curve from the {\it Spitzer} satellite. D16 presented revised properties for the system, and searched for transit timing variations or additional transits which would be caused by the presence of a third body in the system.}

In this work we present extensive simultaneous optical and near-infrared photometry of nine transits of the planet GJ\,1132\,b in front of the star GJ\,1132\,A. We use these data to redetermine the physical properties of the system, improve the fidelity of its orbital ephemeris, and construct a transmission spectrum of the planet. We then interpret the transmission spectrum using suites of theoretical spectra from two model atmosphere codes. We clearly detect the planetary atmosphere but the data in hand are not able to resolve ambiguities in the relative contributions of different molecules to the atmospheric opacity.

At 1.6\Mearth\ GJ\,1132\,b is by a substantial factor the lowest-mass planet with an atmosphere which has been observationally detected. The two other low-mass planets with \reff{claimed detections} are 55\,Cnc\,e \citep{Winn+11apj}, with a mass of $8.08\pm0.31$\Mearth\ \citep{Demory+16nat}, for which \citet{Tsiaras+16apj} found atmospheric features that could most easily be explained by HCN opacity, and GJ\,3470\,b \citep{Bonfils+12aa}, with a mass of $13.7\pm1.6$\Mearth\ \citep{Biddle+14mn}, which shows an enhanced radius in the blue attributable to Rayleigh absorption \citep{Nascimbeni+13aa2,Biddle+14mn,Dragomir+15apj}.

Three other low-mass planets have been subjected to transmission spectroscopy which has failed to reveal atmospheric signatures, most likely due to the presence of clouds \reff{or of a high-metallicity atmosphere with a large mean molecular mass}. They are GJ\,1214\,b \citep{Charbonneau+09nat}, with a mass of $6.26\pm0.91$\Mearth\ \citep{Anglada+13aa}, for which no atmospheric features have been detected \citep{Bean+10nat,Bean+11apj,Demooij+12aa,Kreidberg+14nat}, HD\,97658\,b \citep{Dragomir+13apj}, with a mass of $\er{7.55}{0.83}{0.79}$\Mearth\ \citep{Vangrootel+14apj}, for which there was also a non-detection of the atmosphere \citep{Knutson+14apj2}, and GJ\,436\,b \citep{Gillon+07aa2}, with a mass of $25.4\pm2.1$\Mearth\ \citep{Lanotte+14aa}, for which \citet{Knutson+14nat} found no atmospheric features.

% System                                                                                                         Discovery               Mass-ref
% GJ 1132:    2015Natur.527..204B     Berta-Thompson    M2=0.0051 ± 0.0017  Mjup     M2=1.63 pm 0.54 Mearth  2015Natur.527..204B     this work
% GJ 1214:    xxx12108087             Anglada-Escudé    M2=0.0196 ± 0.0029  Mjup     M2=6.26 pm 0.91 Mearth  2009Natur.462..891C     2013A+A...551A..48A
% HD 97658    2014ApJ...786....2V     Van Grootel       M2=0.0238 +0.0026 -0.0025    M2=7.55 +0.83 -0.79     2013ApJ...772L...2D     2014ApJ...786....2V
% 55 Cnc e    xxx160405725            Demory            M2=0.0254 ± 0.0010           M2=8.08 pm 0.31         2011ApJ...737L..18W     2016Natur.532..207D
% GJ 3470:    2014MNRAS.443.1810B     Biddle            M2=0.0432 ± 0.0051           M2=13.73 pm 1.61        2012A+A...546A..27B     2014MNRAS.443.1810B
% GJ 436:     2014A+A...572A..73L     Lanotte           M2=0.0799 +0.0066 -0.0063    M2=25.4 pm 2.1          2007A+A...472L..13G     2014A+A...572A..73L
% HAT-P-12:   2014ApJ...785..126K     Knutson           M2=0.2089 +0.01 -0.0097

%%%%%%%%%%%%%%%%%%%%%%%%%%%%%%%%%%%%%%%%%%%%%%%%%%%%%%%%%%%%%%%%%%%%%%%%%%%%%%%%%%%%%%%%%%%%%%%%%%%%%%%%%%%%%%%%%%%%%%%%%%%%%%%%%%%%%%%%%%%%%%%%%%%%%

\section{Observations and data reduction}                                                                                             \label{sec:obs}

\begin{figure*} \includegraphics[width=\textwidth,angle=0]{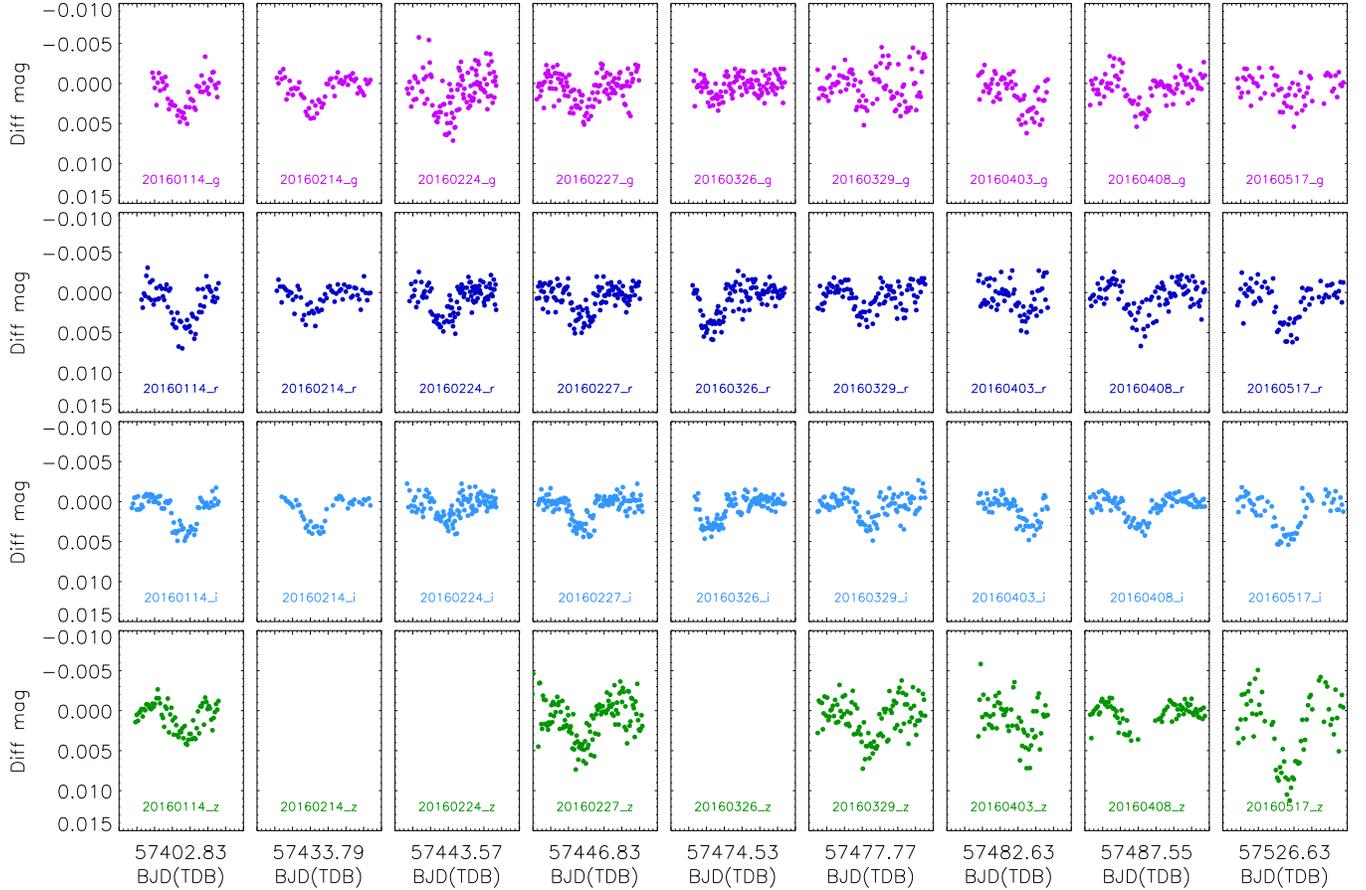}
\caption{\label{fig:plotlc} The optical light curves of GJ\,1132 obtained using GROND,
arranged in rows according to passband and in columns according to date. The filter and
date are encoded in the names of the datafile printed at the base of each panel. Each
plot covers a total of 0.14\,d centred on the time given on the x-axis.} \end{figure*}

\begin{table} \centering
\caption{\label{tab:obslog} Dates and \reff{numbers} of the observations presented in this work.}
\begin{tabular}{lccccccc} \hline
Observing night & \multicolumn{4}{c}{Number of observations} \\
                & $g$ & $r$ & $i$ & $z$ & $J$ & $H$ & $K$ \\
\hline
2016/01/14 &  54 &  63 &  72 &  69 & 252 & 224 &     \\
2016/02/14 &  59 &  59 &  44 &     &     &     &     \\
2016/02/24 & 108 & 104 & 108 &     &     &     &     \\
2016/02/27 & 116 & 121 & 124 & 123 & 372 & 364 & 369 \\
2016/03/26 & 104 & 104 & 101 &     &     &     & 168 \\
2016/03/29 &  94 &  96 &  95 &  97 & 385 & 353 & 358 \\
2016/04/03 &  62 &  63 &  63 &  60 &     &     &     \\
2016/04/08 &  94 &  93 &  94 &  75 & 251 & 251 & 167 \\
2016/05/17 &  62 &  64 &  68 &  65 & 280 & 277 & 273 \\
\hline \end{tabular} \end{table}

Extensive observations of GJ\,1132 were obtained in service mode using the GROND multi-band imager \citep{Greiner+08pasp} mounted on the MPG 2.2\,m telescope at ESO La Silla, Chile. This instrument acquires images simultaneously in four optical and three near-infrared passbands. A total of nine transits were observed, using the telescope-defocussing approach \citep{Alonso+08aa,Me+09mn} with point spread functions (PSFs) of typically 30 pixels in radius. A tenth transit observation suffered from a large systematic effect during transit, due to either weather or instrumental effects, so was not included in our analysis.

The optical bands have response functions similar to those of the SDSS $g$, $r$, $i$ and $z$ filters \citep{Fukugita+96aj}. They are equipped with 2k$\times$2k CCDs which see (approximately) the same 5.4\am$\times$5.4\am\ field, with a plate scale of 0.158\as\,pixel$^{-1}$, and are each read out through two amplifiers. The near-infrared bands have 1k$\times$1k Rockwell HAWAII-1 detectors with a field of view of 10\am$\times$10\am, which fully encompasses the optical field, at a plate scale of 0.6\as\,pixel$^{-1}$

Several phenomena affected the quality of our optical observations. Firstly, GJ\,1132 is observationally difficult for the GROND imager due to its low \Teff\ and concomitant large variations in flux level through the optical wavelength region. The requirement for a common exposure time and focus level meant a compromise had to be made between obtaining adequate count rates in the $g$ band whilst avoiding overexposure in the $z$ band. Three of the $z$-band observing sequences suffered from saturation effects which precluded the extraction of reliable photometry from these images. Secondly, an issue was encountered in the $r$ and $z$ bands attributable to electronic noise in the CCD controllers. This caused approximately 0.3\% of the pixels on each image to be assigned count rates near the bias level, with the identities of the affected pixels varying at random with each image. When these coincided with the point spread functions (PSFs) of the target or comparison stars they caused a non-astrophysical drop in the number of counts detected from the star, increasing the scatter on the photometry. A non-repeating effect such as this cannot be calibrated out, but a partial mitigation of the effect was achieved by detecting each pixel with anomalously low count rates and replacing its value with the mean level from the adjacent eight pixels. Thirdly, the small field of view of GROND meant that the $r$, $i$ and $z$ bands suffered from a lack of decent comparison stars. Photometry was therefore performed against comparison stars which were significantly fainter than GJ\,1132 itself. A brief observing log of the nine included transits is given in Table\,\ref{tab:obslog}.

The optical data were reduced using the {\sc defot} pipeline \citep{Me+09mn,Me+14mn}, which performs aperture photometry using the {\sc aper} routine from {\sc daophot} \citep{Stetson87pasp} contained in the NASA {\sc astrolib} library\footnote{{\tt http://idlastro.gsfc.nasa.gov/}}. The aperture positions and sizes were specified manually in order to yield data with the lowest statistical and systematic errors. The science images were not calibrated using bias or flat-field frames as these tended to have little effect on the final light curves beyond a slight increase in the scatter of the datapoints. Differential magnitudes for each light curve were formed versus an optimal ensemble of comparison stars, whilst simultaneously fitting for and removing a quadratic trend of magnitude with time caused primarily by airmass variations. The timestamps were converted into the BJD(TDB) system using routines from \citet{Eastman++10pasp}. As a final step, we rescaled the errorbars for each light curve to obtain a reduced $\chi^2$ of $\chir=1.0$ versus a fitted model (see below). The light curves are shown in Fig.\,\ref{fig:plotlc} and will be made available at the CDS\footnote{{\tt http://cdsweb.u-strasbg.fr}}.

The infrared data were reduced using standard IDL routines following the methods outlined by \citet{Chen+14aa}. In brief, we constructed a master dark frame and a master flat field by median-combining the individual dark and sky flat field images obtained before the science observations. From each science image we subtracted the master dark frame and divided by the normalised master flat field. \reff{No correction for non-linearity was applied to the data because the counts were below the level at which linearity becomes important for the GROND infrared detectors.} We obtained the light curves using aperture photometry routines to extract the flux of the target and several comparison stars. We also tried to correct for the odd-even readout pattern present along the x-axis, but found no improvement in the light curve so decided not to include the correction for this effect when obtaining our final light curves. We were able to obtain useful results for five transits in each band (Table\,\ref{tab:obslog}); the remaining datasets suffered from correlated noise features which were larger than the transit depth.

%%%%%%%%%%%%%%%%%%%%%%%%%%%%%%%%%%%%%%%%%%%%%%%%%%%%%%%%%%%%%%%%%%%%%%%%%%%%%%%%%%%%%%%%%%%%%%%%%%%%%%%%%%%%%%%%%%%%%%%%%%%%%%%%%%%%%%%%%%%%%%%%%%%%%

\section{Light curve modelling}                                                                                                        \label{sec:lc}

\begin{figure*} \includegraphics[width=\textwidth,angle=0]{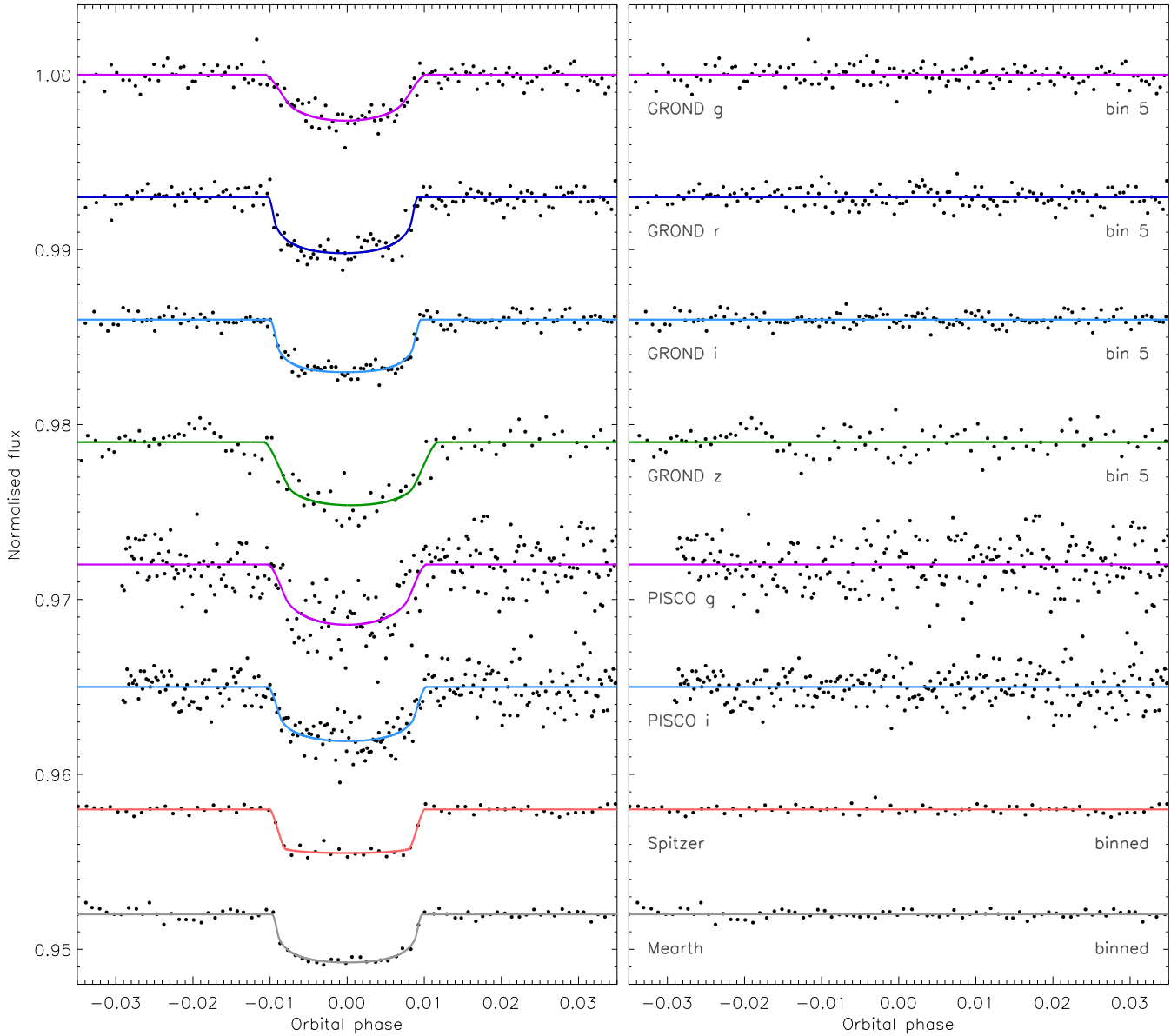}
\caption{\label{fig:lcfit} Optical light curves of GJ\,1132 from this work (GROND),
BT15 (PISCO) \reff{and D16 ({\it Spitzer} and MEarth)}, compared to the {\sc jktebop} best
fits.The GROND data have been binned by a factor of \reff{five} and plotted versus orbital phase
in order to make the plot clearer; all analysis in this work was based on the original
data. \reff{The {\it Spitzer} and MEarth datasets are digitized versions of data heavily binned before
plotting.} The residuals of the fits are plotted at the base of the figure, offset
from unity. Labels give the source and passband for each dataset. The polynomial
baseline functions have been removed from the data before plotting.} \end{figure*}

\begin{table*} \centering \caption{\label{tab:lcfit} Parameters of the fit to the
GROND and PISCO light curves of GJ\,1132 from the {\sc jktebop} analysis.}
\begin{tabular}{l r@{\,$\pm$\,}l r@{\,$\pm$\,}l r@{\,$\pm$\,}l r@{\,$\pm$\,}l r@{\,$\pm$\,}l}
\hline
Source & \mc{$r_{\rm A}+r_{\rm b}$} & \mc{$k$} & \mc{$i$ ($^\circ$)} & \mc{$r_{\rm A}$} & \mc{$r_{\rm b}$} \\
\hline
GROND $g$-band            & 0.102 & 0.028 & 0.0535 & 0.0060 & 85.4 & 2.0 & 0.097 & 0.026 & 0.0052 & 0.0019 \\
GROND $r$-band            & 0.066 & 0.019 & 0.0517 & 0.0040 & 88.5 & 2.1 & 0.063 & 0.018 & 0.0032 & 0.0012 \\
GROND $i$-band            & 0.070 & 0.014 & 0.0514 & 0.0023 & 87.9 & 1.6 & 0.067 & 0.013 & 0.0034 & 0.0008 \\
GROND $z$-band            & 0.106 & 0.028 & 0.0611 & 0.0061 & 85.3 & 1.9 & 0.100 & 0.026 & 0.0061 & 0.0021 \\
PISCO $g$-band            & 0.117 & 0.039 & 0.0633 & 0.0086 & 84.5 & 2.6 & 0.110 & 0.036 & 0.0070 & 0.0029 \\
PISCO $i$-band            & 0.086 & 0.030 & 0.0541 & 0.0050 & 86.6 & 2.4 & 0.082 & 0.028 & 0.0044 & 0.0019 \\
{\it Spitzer} 4.5\,$\mu$m & \reff{0.090} & \reff{0.014} & \reff{0.0496} & \reff{0.0010} & \reff{86.3} & \reff{1.0} & \reff{0.086} & \reff{0.013} & \reff{0.0043} & \reff{0.0007} \\
MEarth                    & \reff{0.064} & \reff{0.024} & \reff{0.0492} & \reff{0.0026} & \reff{88.5} & \reff{1.8} & \reff{0.061} & \reff{0.024} & \reff{0.0030} & \reff{0.0013} \\
\hline
Final results             & \reff{0.0814} & \reff{0.0072} & \reff{0.05041} & \reff{0.00086} & \reff{86.58} & \reff{0.63} & \reff{0.0775} & \reff{0.0068} & \reff{0.00397} & \reff{0.00042} \\
\hline \end{tabular} \end{table*}

The optical GROND data were combined into four sets, one for each of the $g$, $r$, $i$ and $z$ passbands, and then each was modelled using the {\sc jktebop} code \citep{Me13aa}. We fitted for the orbital inclination ($i$), time of mid-transit ($T_0$), and the sum and ratio of the fractional radii ($r_{\rm A}+r_{\rm b}$ and $k=\frac{r_{\rm b}}{r_{\rm A}}$) where the fractional radii are those of the star and planet in units of the orbital semimajor axis ($r_{\rm A,b}=\frac{R_{\rm A,b}}{a}$). The orbital period was fixed to the known value (see Section\,\ref{sec:tmin}) and the orbit was assumed to be circular. A quadratic function versus time was applied to the magnitude values for each observed transit in order to propagate the uncertainty in this from the light curve generation process.

Limb darkening was incorporated using each of five laws \citep[see][]{Me08mn} and with the nonlinear coefficient fixed. Fits were obtained with the linear coefficient either fixed or fitted in order to see how this changed the results. We adopted limb darkening coefficients calculated using the {\sc phoenix} model atmospheres by \citet{Claret04aa2}, and linearly interpolated the coefficients to the stellar temperature of $\Teff=3270\pm140$\,K (BT15).

% Nearby star RA = 10 14 49.75 and Dec = -47 09 24.4 from the DSS2 red plate. Separation from GJ 1132 is 6.5 arcsec from the GROND images.

An additional complication arises due to the presence of the faint nearby star USNO B1.0 0428-0265237 \citep{Monet+03aj}, which was separated from GJ\,1132 by approximately 6.5\as\ at the times of our observations. A small fraction of the flux from this star leaked into the aperture used to measure the counts of GJ\,1132. From a simple PSF model we measured the amount of contamination as $1.2\pm0.3$\% in the $g$-band, $0.5\pm0.2$\% in $r$, $0.3\pm0.1$\% in $i$ and $0.2\pm0.1$\% in $z$, where the errorbars are very conservative. This effect was included as `third light' in the {\sc jktebop} model following the approach in \citet{Me10mn}. The best fits are plotted in Fig.\,\ref{fig:lcfit} in the case of fixed limb darkening coefficients.

We also modelled the two best light curves of GJ\,1132 from BT15, which are the $g$- and $i$-band data from the PISCO imager. Their appearance in Fig.\,\ref{fig:lcfit} differs from that in BT15, leading us to investigate the discrepancy. We checked for correlation with the instrumental parameters supplied with the flux measurements (airmass, x-position, y-position, PSF width and sky background level), finding linear Pearson correlation coefficients less than 0.13 in all cases. We therefore conclude that the difference in appearance is \reff{only} because BT15 binned their data into 1.5\,min intervals before plotting it.

\reff{Shortly before our own manuscript was submitted, D16 presented an analysis of GJ\,1132 based on observations of 21 transits with the MEarth telescopes and 100\,hr with the {\it Spitzer} satellite using the 4.5\,$\mu$m channel of the IRAC imager. Whilst the original data are not available to us, we have been able to obtain binned light curves from both facilities by digitizing fig.\,2 in D16. We used the scatter of the data around the best-fit model to assess the photometric precisions of the datasets. For the {\it Spitzer} data we used limb darkening coefficients from \citet{Claret++12aa} and for the MEarth data we adopted the mean of the coefficients for the $i$ and $z$ bands used above. For the {\it Spitzer} light curve we found that fitting for any limb darkening coefficients returned a nett {\em limb brightening}, due to the scatter of the data during transit. The MEarth light curve was able to support the fitting of one limb darkening coefficient. Results for both datasets are included in Table\,\ref{tab:lcfit}.}

Table\,\ref{tab:lcfit} gives the best-fitting photometric parameters for each light curve, together with the weighted mean value for each parameter. Error estimates for the photometric parameters were obtained using Monte Carlo \citep{Me++04mn2} and residual-permutation simulations \citep{Me08mn}, and the larger value retained in each case. These were then inflated to account for any variation between the solutions adopting the five different limb darkening laws \reff{by adding in quadrature the largest difference in values of each parameter from solutions with the various limb darkening laws}. Finally, the weighted mean values agree well for all parameters, indicating that the errorbars are robust. \reff{In all cases we have chosen to adopt the solutions with all limb darkening coefficients fixed. We find that the alternative solutions with one fitted coefficient agree to well within the uncertainties, with the exception of the {\it Spitzer} data as discussed above.}

\reff{There is a significant disagreement between the photometric parameters determined by ourselves and those by BT15 and D16. An inspection of Table\,\ref{tab:lcfit} shows that there is a correlation between the parameters $r_{\rm A}$ and $i$, whereby higher-inclination solutions give a smaller fractional radius for the star. BT15 and D16 obtain solutions with higher $i$ than our own, resulting in lower $r_{\rm A}$ values of $0.0625\pm0.0043$ and $0.0605^{+0.0027}_{-0.0022}$, respectively. This can be explained in the case of BT15 because $r_{\rm A}$ is a proxy for the density of the star \citep{Sozzetti+07apj,Me17pasp}, so this quantity was effectively fixed in the light curve solutions by the imposition of external constraints on both the mass and radius of the host star.}

\reff{Such a constraint was not imposed in the current work or D16, so the discrepancy between these two works remains. We have checked and ruled out the possibility that it could be caused by the treatment of limb darkening or third light, by quantifying the effect of varying our own treatment of these phenomena within reasonable limits. We also tried numerically integrating the fitted model \citep{Me11mn} to match the cadence of the {\it Spitzer} and MEarth data but found this caused little change in the solutions. We also tried various initial values for the fitted parameters, including the values found by D16, but the fits to data converged on the same solutions as given in Table\,\ref{tab:lcfit}. Solution determinacy is therefore not an issue, and the cause of the discrepancy remains unidentified. It is most likely due to differences in the overall analysis procedure.}

%%%%%%%%%%%%%%%%%%%%%%%%%%%%%%%%%%%%%%%%%%%%%%%%%%%%%%%%%%%%%%%%%%%%%%%%%%%%%%%%%%%%%%%%%%%%%%%%%%%%%%%%%%%%%%%%%%%%%%%%%%%%%%%%%%%%%%%%%%%%%%%%%%%%%

\section{Orbital ephemeris}                                                                                                          \label{sec:tmin}

\begin{table} \centering
\caption{\label{tab:tmin} Times of mid-transit from the data presented in this work.}
\begin{tabular}{rr@{\,$\pm$\,}l} \hline
\reff{Orbital cycle} & \multicolumn{2}{c}{Transit time (BJD/TDB)} \\
\hline
  0.0 & 2457184.55786 & 0.00032 \\
134.0 & 2457402.83351 & 0.00026 \\
153.0 & 2457433.78361 & 0.00024 \\
159.0 & 2457443.55662 & 0.00039 \\
161.0 & 2457446.81540 & 0.00028 \\
178.0 & 2457474.50661 & 0.00030 \\
180.0 & 2457477.76559 & 0.00043 \\
183.0 & 2457482.65244 & 0.00046 \\
186.0 & 2457487.53908 & 0.00031 \\
210.0 & 2457526.63273 & 0.00030 \\
\hline \end{tabular} \end{table}

Many of our light curves have a relatively high scatter compared to the transit depth. We therefore combined all four light curves of each individual transit into a single dataset and fitted them with {\sc jktebop} whilst fixing the sum of the radii ($r_{\rm A} + r_{\rm b}$) to the best fit from Table\,\ref{tab:lcfit}. This yielded nine measured times of mid-transit, which are given in Table\,\ref{tab:tmin}.

We augmented these timing measurements with the reference time of mid-transit given by BT15 from a global analysis of their photometric data ($2457184.55786\pm0.00032$) \reff{and with the 22 timings given by D16 from orbital cycle 91 onwards}. The timings were then fit with a straight line to yield the linear ephemeris:
$$ T_0={\rm BJD(TDB)} \,\, 2457184.55759 (30) \, + \, 1.6289287 (18) \times E $$
where the fit has $\chir=1.13$ and the uncertainties (given in brackets and relative to the preceding digit) have been increased \reff{by $\sqrt{1.13}$} to account for this. The scatter around the best fit gives no indication of deviations from a linear ephemeris, so we do not find any evidence for transit timing variations.

%%%%%%%%%%%%%%%%%%%%%%%%%%%%%%%%%%%%%%%%%%%%%%%%%%%%%%%%%%%%%%%%%%%%%%%%%%%%%%%%%%%%%%%%%%%%%%%%%%%%%%%%%%%%%%%%%%%%%%%%%%%%%%%%%%%%%%%%%%%%%%%%%%%%%

\section{Physical properties of GJ\,1132}                                                                                          \label{sec:absdim}

\begin{table*} \centering \caption{\label{tab:model} Derived physical properties of GJ\,1132
from the current work. The values found by BT15 \reff{and D16} are included for comparison.}
\begin{tabular}{l l l c c c} \hline
Quantity                & Symbol           & Unit         & \reff{This work}        & BT15            & \reff{D16}                  \\
\hline
Stellar mass            & $M_{\rm A}$      & \Msun        & $0.181\pm0.019$         & $0.181\pm0.019$ &                             \\
Stellar radius          & $R_{\rm A}$      & \Rsun        & $0.255\pm0.023$         & $0.207\pm0.016$ & \er{0.2105}{0.0102}{0.0085} \\
Stellar surface gravity & $\log g_{\rm A}$ & \,cgs        & $4.881\pm0.074$         &                 &                             \\
Stellar density         & $\rho_{\rm A}$   & \psun        & \er{10.9}{3.4}{2.4}     & $21.0\pm4.3$    & \er{19.4}{2.6}{2.5}         \\
Planet mass             & $M_{\rm b}$      & \Mearth      & $1.63\pm0.54$           & $1.62\pm0.55$   &                             \\
Planet radius           & $R_{\rm b}$      & \Rearth      & $1.43\pm0.16$\,$^\star$ & $1.16\pm0.11$   & $1.130\pm0.056$             \\
Planet surface gravity  & $g_{\rm b}$      & \mss         & \er{7.8}{3.4}{2.8}      & $11.7\pm4.3$    &                             \\
Planet density          & $\rho_{\rm b}$   & g\,cm$^{-3}$ & \er{3.1}{1.7}{1.2}      & $6.0\pm2.5$     & $6.2\pm2.0$                 \\
Equilibrium temperature & \Teq\            & K            & $644\pm38$              & $579\pm15$      &                             \\
Orbital semimajor axis  & $a$              & au           & $0.01533\pm0.00053$     &                 &                             \\
\hline \end{tabular}
\begin{flushleft}
\reff{$^\star$\,This value of $R_b$ is averaged over all observed passbands. However, while interpreting the spectrum
we consider the $z$-band radius at $1.57\pm0.05$\Rearth\ to be contributed by the planetary atmosphere whereas the
remaining datapoints are consistent with a continuum, the ``surface'' of the planet, at a bulk radius of 1.375\Rearth.}
\end{flushleft}
\end{table*}

The physical properties of the GJ\,1132 system were established by BT15 using the following steps. Firstly, the measured trigonometric parallax (BT15, \citealt{Jao+05aj}) and 2MASS $JHK$ magnitudes were used to determine the $K$-band absolute \reff{magnitude} of the star, from which its mass was found via the calibration by \citet{Delfosse+00aa}. Secondly, an empirical \reff{calibration of mass versus density} \citep{Hartman+15aj} was used to find the density and thus radius of the star. Thirdly, the transit light curves were fitted with the stellar density constrained to the value found in \reff{the second step}, which is equivalent to constraining $r_{\rm A}$.

We are in a position to modify this approach to rely less on general empirical calibrations and more on data obtained for GJ\,1132 itself. We adopted the stellar mass of $0.181\pm0.019$\Msun\ from BT15, and used this along with the photometric parameters measured in Section\,\ref{sec:lc} to determine the physical properties of the system using standard equations \citep[e.g.][]{Hilditch01book}. \reff{This process in effect used $r_{\rm A}$ as a proxy for the density of the star \citep{SeagerMallen03apj}, which combined with its mass yields its radius. The properties of the planet could then be determined relative to those of the star using $k$ and the velocity amplitude of the star measured by BT15, $2.76\pm0.92$\ms}. The uncertainties were propagated by a Monte Carlo approach.

\reff{The measured physical properties of the GJ\,1132 system are} summarised in Table\,\ref{tab:model}. It can be seen that we find a rather lower density for the star -- measured directly from the light curves whereas BT15 obtained their value from a calibration of stellar properties -- which results in increased radii for both star and planet. The measured density of GJ\,1132\,b has decreased by a factor of two (1.8$\sigma$), which has a significant impact on the likely properties of this body.

%%%%%%%%%%%%%%%%%%%%%%%%%%%%%%%%%%%%%%%%%%%%%%%%%%%%%%%%%%%%%%%%%%%%%%%%%%%%%%%%%%%%%%%%%%%%%%%%%%%%%%%%%%%%%%%%%%%%%%%%%%%%%%%%%%%%%%%%%%%%%%%%%%%%%

\section{Transmission spectrum of GJ\,1132\,b}                                                                                  \label{sec:transspec}

\begin{figure} \includegraphics[width=\columnwidth,angle=0]{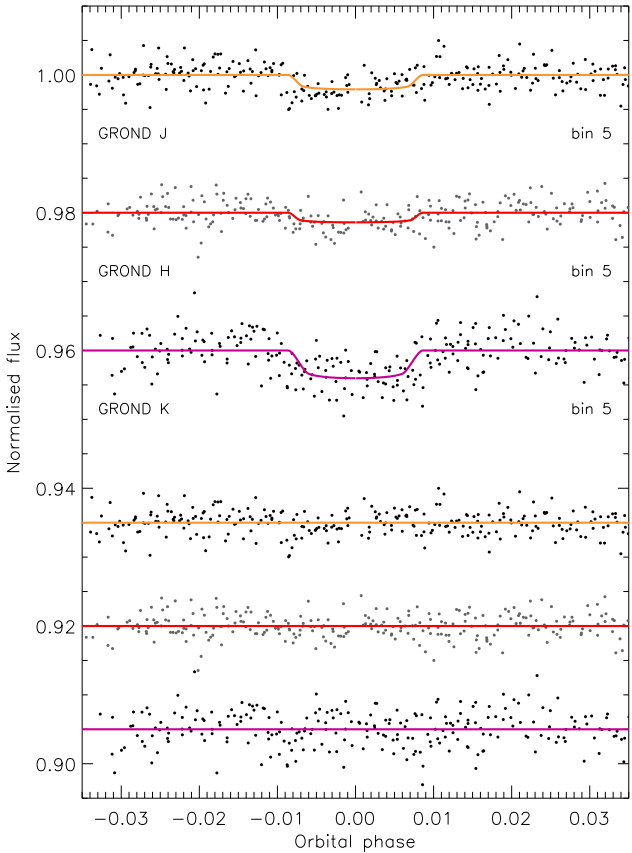}
\caption{\label{fig:lcfitJHK} Near-IR light curves of GJ\,1132 compared to the {\sc jktebop}
best fits. The data have been binned by a factor of \reff{five} and plotted versus orbital phase in
order to make the plot clearer. Labels give the source and passband for each dataset. The
polynomial baseline functions have been removed from the data before plotting. The residuals
of the fits are plotted at the base of the figure, offset from unity.} \end{figure}

\begin{table} \centering
\caption{\label{tab:rb} Values of $r_{\rm b}$ for each of the six light curves. The errorbars
in this table exclude all common sources of uncertainty so should only be used to interpret
relative differences in $r_{\rm b}$. To convert $r_{\rm b}$ to $R_{\rm b}$ we multiplied by
23455.0, the orbital semimajor axis in units of \Rearth. The central wavelengths and full
widths at half maximum transmission are given for the GROND filters.}
\begin{tabular}{lccr@{\,$\pm$\,}l} \hline
Passband & Central         & Band full  & \mc{$r_{\rm b}$} \\
         & wavelength (nm) & width (nm) & \mc{ }           \\
\hline
GROND $g$ & 477   & 138   & \reff{0.00382} & \reff{0.00011} \\
GROND $r$ & 623   & 138   & \reff{0.00402} & \reff{0.00009} \\
GROND $i$ & 763   & 154   & \reff{0.00386} & \reff{0.00006} \\
GROND $z$ & 913   & 137   & \reff{0.00446} & \reff{0.00015} \\
PISCO $g$ &       &       & \reff{0.00438} & \reff{0.00010} \\
PISCO $i$ &       &       & \reff{0.00396} & \reff{0.00007} \\
GROND $J$ & 1230  & 410   & \reff{0.00354} & \reff{0.00045} \\
GROND $H$ & 1645  & 420   & \reff{0.00324} & \reff{0.00044} \\
GROND $K$ & 2165  & 570   & \reff{0.00473} & \reff{0.00058} \\
% {\it Spitzer} & \reff{4485} & \reff{530} & \reff{0.00385} & \reff{0.00008} \\
\hline \end{tabular} \end{table}

The main aim of the current work is to explore if constraints on the atmospheric composition of GJ\,1132\,b can be obtained using multi-band photometry. We therefore sought to determine the radius of the planet in each of the passbands for which we possess a light curve. It is important to remove sources of uncertainty common to all passbands, so that the significance of any {\em relative} variations between passbands can be assessed.

Following the approach of \citet{Me+12mn2}, we modelled the nine available light curves (GROND $griz$, GROND $JHK$ and PISCO $gi$) with $r_{\rm A}$, $i$ and the orbital period fixed to the final values determined above. We included as fitted parameters $r_{\rm b}$, the reference time of mid-transit, and the quadratic function of magnitude versus time for each transit. We adopted quadratic limb darkening with the linear coefficient fitted and the nonlinear coefficient fixed to the values used above. \reff{As the $r_{\rm A}$ and $i$ parameters are fixed, it is not only more tractable but also more important to fit for limb darkening in order to propagate its uncertainty and avoid biases arising from the use of theoretical values.} Third light was included for the optical bands and constrained as in Section\,\ref{sec:lc}. For the $JHK$ bands we neglected third light, as it is negligible at infrared wavelengths, but iteratively clipped points lying more than 3$\sigma$ from the best fit in order to remove scattered data which was biasing the fitting process. The \reff{near-IR} light curves and best fits are shown in Fig.\,\ref{fig:lcfitJHK}. \reff{We did not include the data from D16 in this analysis because imperfections in the digitization process could significantly affect our results. In addition, the wavelength resolution of the MEarth data is poor and the {\it Spitzer} data cannot support the fitting of limb darkening.}

\begin{figure} \includegraphics[width=\columnwidth,angle=0]{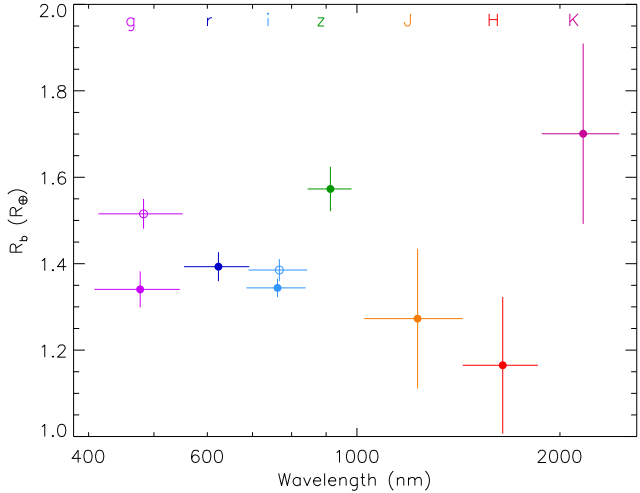}
\caption{\label{fig:rvary} Transmission spectrum of GJ\,1132\,b: the measured planetary radius
($R_{\rm b}$) as a function of the central wavelength of the passbands used. The passband names
are given at the top of the plot. The horizontal lines indicate the FWHM of the passband used and
the vertical lines show the {\rm relative} errorbars in the $R_{\rm b}$ measurements. \reff{Filled
circles show results from GROND data and open circles those from PISCO data.}} \end{figure}

For each light curve we determined the best-fitting $r_{\rm b}$, and obtained its uncertainty via \reff{1000} Monte Carlo simulations. The results of this process are given in Table\,\ref{tab:rb} along with details of the passbands used, and shown in Fig.\,\ref{fig:rvary}. For the purposes of visualising our results we assumed that the PISCO passbands correspond to those from \citet{Fukugita+96aj}. It can be seen that the $g$, $r$ and $i$ passbands agree well, except for a discrepancy between our own $g$-band results and those from the PISCO $g$-band data from BT15. We suggest that our $r_{\rm b}$ value is more reliable as it is based on observations of nine transits whereas the PISCO data cover only one transit. It is also apparent that the radius of the planet in $z$ is significantly larger than that in the other passbands, the discrepancy being a maximum of 4.1$\sigma$ versus our $i$-band value of $r_{\rm b}$.

\subsection{Testing the robustness of the results}

\begin{figure} \includegraphics[width=\columnwidth,angle=0]{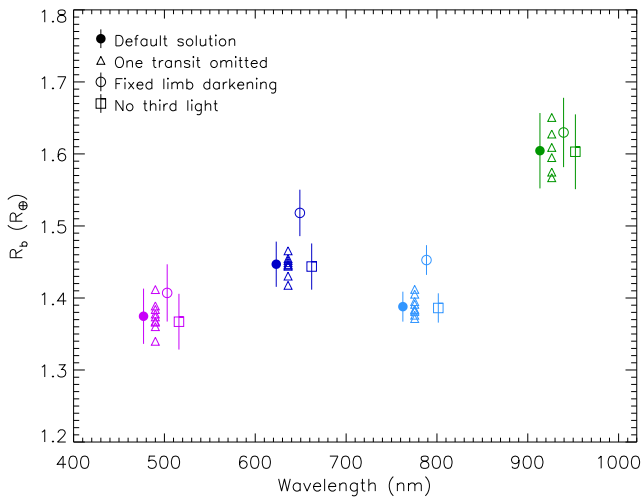}
\caption{\label{fig:rvary:test} Results of testing the measurements of the planetary radius
($R_{\rm b}$) in the GROND $griz$ passbands. The vertical lines show the {\rm relative}
errorbars in the $R_{\rm b}$ measurements. This figure shows the different measurements
of $R_{\rm b}$ obtained whilst performing several tests on the robustness of the results.
The results from the various tests are offset in wavelength for clarity.} \end{figure}

The optical transmission spectrum is the most important result of the current work, so we have deployed a suite of tests to probe the reliability of the radius values and errorbars found for each passband. These have been applied to the optical bands only, as it is here that the measurements are most precise and extensive, and because the optical bands are the most affected by issues such as limb darkening and contaminating light.

\subsubsection{Sensitivity to individual transit observations}

\begin{figure} \includegraphics[width=\columnwidth,angle=0]{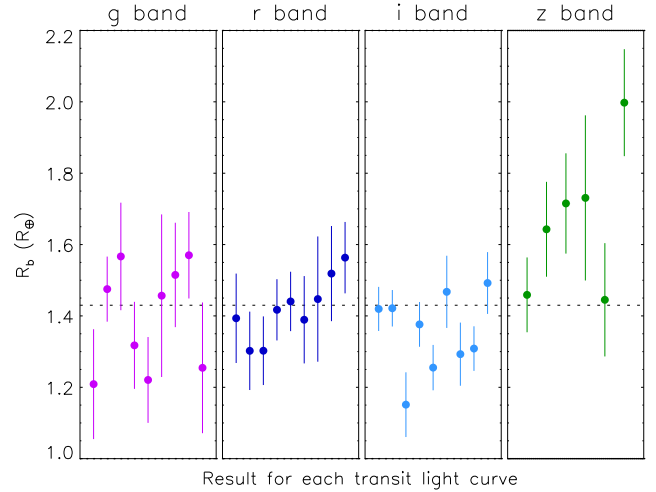}
\caption{\label{fig:rvary:indiv} \reff{Measured planetary radius for each individual
transit in the GROND $griz$ bands, with relative errorbars. The horizontal line shows
the overall radius measurements obtained in Section\,\ref{sec:absdim}.}} \end{figure}

\begin{table*} \centering
\caption{\label{tab:rvary:indiv} \reff{Measured planetary radius for each individual transit
in the GROND $griz$ bands, with relative errorbars, as plotted in Fig.\,\ref{fig:rvary:indiv}.}}
\begin{tabular}{lr@{\,$\pm$\,}lr@{\,$\pm$\,}lr@{\,$\pm$\,}lr@{\,$\pm$\,}l} \hline
Observing night & \multicolumn{8}{c}{Measured planetary radius (\Rearth)} \\
                & \mc{$g$} & \mc{$r$} & \mc{$i$} & \mc{$z$} \\
\hline
2016/01/14 & 1.209 & 0.154 & 1.393 & 0.125 & 1.420 & 0.062 & 1.459 & 0.105 \\
2016/02/14 & 1.475 & 0.091 & 1.302 & 0.110 & 1.422 & 0.051 \\
2016/02/24 & 1.567 & 0.151 & 1.302 & 0.096 & 1.151 & 0.091 \\
2016/02/27 & 1.318 & 0.122 & 1.417 & 0.086 & 1.376 & 0.063 & 1.643 & 0.133 \\
2016/03/26 & 1.221 & 0.120 & 1.441 & 0.083 & 1.255 & 0.064 \\
2016/03/29 & 1.457 & 0.228 & 1.389 & 0.122 & 1.468 & 0.101 & 1.715 & 0.141 \\
2016/04/03 & 1.515 & 0.146 & 1.447 & 0.175 & 1.293 & 0.089 & 1.731 & 0.231 \\
2016/04/08 & 1.570 & 0.121 & 1.519 & 0.133 & 1.309 & 0.063 & 1.445 & 0.159 \\
2016/05/17 & 1.255 & 0.183 & 1.563 & 0.100 & 1.492 & 0.087 & 1.998 & 0.150 \\
\hline \end{tabular} \end{table*}

A cursory inspection of Fig.\,\ref{fig:plotlc} reveals that the transit depth in each passband may vary between different transits. The most obvious indicators are an unusually shallow transit in $g$ on 2016/03/26 and an apparently very deep transit in $z$ on 2016/05/17. This possibility can be probed by determining a set of solutions in each passband, each one based on the full dataset but for one transit omitted. We performed this analysis for the GROND data and plot the results in Fig.\,\ref{fig:rvary:test}. The PISCO data were not considered: they are unsuitable for this analysis because they cover only one transit.

We found that the planetary radius measurements were not significantly affected by the omission of individual transit datasets: all results exist within the 1$\sigma$ errorbars found for our default solution. We conclude that our results are robust against the omission of individual light curves. This is a particularly powerful test in the current case, because of the large number of datasets obtained in each band, and makes us confident in the reliability of both the measured values and errorbars of the optical transmission spectrum.

\reff{We also modelled every optical transit individually and show the resulting $R_{\rm b}$ values in Fig.\,\ref{fig:rvary:indiv}. The $z$-band values of $R_{\rm b}$ show a clear increase relative to those for other bands, peaking at a 4.4$\sigma$ difference between $i$ and $z$. Even if we reject the highest radius measurement, the last one in the $z$ band, we obtain a difference in $R_{\rm b}$ between the $i$ and $z$ bands significant at the 3.1$\sigma$ level. The measurements exhibited in Fig.\,\ref{fig:rvary:indiv} are tabulated in Table\,\ref{tab:rvary:indiv}.}

\subsubsection{Treatment of limb darkening}

\begin{figure*} \includegraphics[width=\textwidth,angle=0]{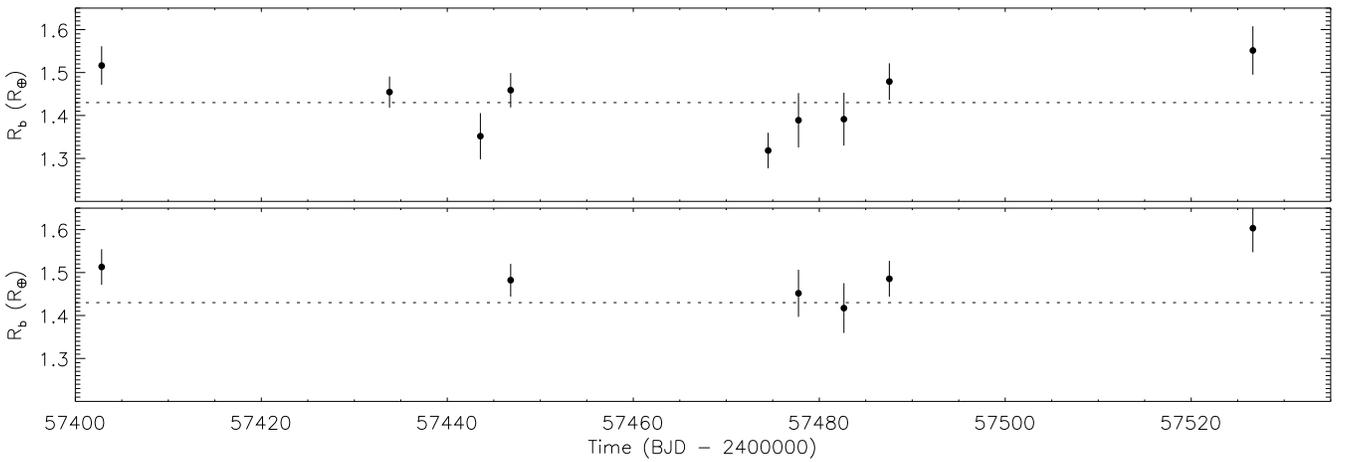}
\caption{\label{fig:timevary} Planetary radius measurements from individual transits.
The upper panel shows results for fitting the $g$, $r$ and $i$-band data together.
The lower panel includes also the $z$-band data, which are available for six of the
nine transits. The errorbars include only those contributions specific to individual
transits, and do not include sources of uncertainty common to all transits. The dotted
line shows the overall measured value of $R_{\rm b}=1.43$\Rearth\ found in
Section\,\ref{sec:absdim}.} \end{figure*}

Our default approach to obtaining the transmission spectrum was to model limb darkening using the quadratic law, with the linear coefficient included as a parameter of the fit and the nonlinear coefficient fixed to a value interpolated from theoretical predictions by \citet{Claret04aa2}. This causes a dependence on theoretical model atmospheres, which is an important consideration for a planet host star as cool as GJ\,1132\,A. We therefore calculated the planetary radius measurements in the GROND and PISCO bands using an alternate approach: fixing both limb darkening coefficients.

Fig.\,\ref{fig:rvary:test} shows the effect on the results for the GROND bands; the PISCO bands show a similar effect. We find that fixing both limb darkening coefficients causes the solution to move to higher planetary radius. Whilst the effect is similar for all four bands, it affects $r$ and $i$ more than $g$ and $z$. It therefore modifies the transmission spectrum at the 1--2$\sigma$ level, in particular the slope seen through the optical wavelength region. As the $g$ and $z$ bands are relatively unaffected, limb darkening cannot explain the anomalously large radius we find in the $z$ band. However, we conclude that observed transmission spectra can be significantly affected by the treatment of limb darkening and urge future studies to check for this possibility in all cases.

\subsubsection{Effect of contaminating light}

Our default solution includes a third-light contribution from a nearby star. This phenomenon can have a significant effect on the transmission spectrum when the spectral energy distribution of the target and contaminant differ \citep{MeEvans16mn}, as is the case here. We tested the importance of this effect by calculating solutions for the GROND data whilst neglecting the third-light contribution.

Once again, the results are represented in Fig.\,\ref{fig:rvary:test} and show that the contamination from the nearby star has a negligible effect on the results. The effect is $0.12\sigma$ for the $g$ band and smaller for the other bands. The PISCO data will be similarly (un)affected. We conclude that our transmission spectrum measurements are robust against an amount of contaminating light significantly in excess of that currently known for GJ\,1132.

\subsubsection{Temporal variability}

We modelled the light curves from all passbands but for each transit individually, in order to check for the presence of temporal variability in the planet. To deal with the complication of having $z$-band data available for only a subset of the transits, we obtained results for the $g$, $r$ and $i$ bands together, and also for all four optical bands when possible. \reff{Due to the scatter in the data we fixed the limb darkening coefficients for this analysis, using the linear coefficients 0.39 for $gri$ and 0.31 for $griz$, and the quadratic coefficients 0.45 for $gri$ and 0.51 for $griz$.}

Our results are shown in Fig.\,\ref{fig:timevary}. %We find several outliers on this plot, and these are generally associated with an increased scatter in the transit light curves.
There is no overall trend in measured planet radius during our observations. %Outliers such as those seen in Fig.\,\ref{fig:timevary} could arise from dark spots on the surface of the star, although this is unlikely in the current case

\subsubsection{Stellar activity}

Low-mass stars frequently show dark starspots, which are capable of modifying the transit depth. Starspots occulted by the planet will make the transit shallower, and unocculted starspots will make it deeper \citep[e.g.][]{Ballerini+12aa,Tregloan++13mn,Oshagh+13aa}. GJ\,1132\,A is a relatively old star so will not show strong activity, but the fact that a rotation period has been observed ($125$\,d, BT15) means that it does show {\em some} starspots. However, there is evidence that many M dwarfs show a large number of small spots \citep{JacksonJeffries12mn,JacksonJeffries13mn}, which would greatly reduce the effect of spot activity because the occulted part of the stellar surface would be very similar to the unocculted areas. This is consistent with the lack of starspot features seen in transits of planets in front of M dwarfs \citep[e.g.][]{Mancini+14aa,Awiphan+16mn}.

The very small radius of GJ\,1132\,b to that of GJ\,1132\,A means that unocculted starspots would be the dominant contributor to transit depth variations for this system. Because starspots are cooler than the rest of the stellar surface, unocculted spots would have a wavelength-dependent effect on the transit depth. They would cause transits to become gradually deeper as one observed at bluer wavelengths. This does not provide an explanation for the current case, where the $z$-band radius is significantly larger than the $gri$-band points. For completeness, we note that unocculted plage could have the opposite effect \citep[see also][]{Oshagh+14aa}, but this has never been observed. It would also require the plage to have a very clumpy distribution \reff{in order to not be occulted during the transits we observed}, which is not the situation \reff{seen} for the Sun, and would not cause an abrupt change in transit depth as found in the $z$-band.

Finally, we note that our observations of GJ\,1132 were distributed over 124\,d, and therefore fortuitously sample the 125\,d rotation period of GJ\,1132\,A very well. Moreover, observations in the different passbands were obtained simultaneously so temporal variations are unable to affect the relative transit depth measurements. We therefore conclude that our observations are not significantly affected by spot activity in the host star. Having checked for and ruled out issues due to individual transits, contaminating light, planetary and stellar variability, we conclude that our transmission spectrum is robust.

%%%%%%%%%%%%%%%%%%%%%%%%%%%%%%%%%%%%%%%%%%%%%%%%%%%%%%%%%%%%%%%%%%%%%%%%%%%%%%%%%%%%%%%%%%%%%%%%%%%%%%%%%%%%%%%%%%%%%%%%%%%%%%%%%%%%%%%%%%%%%%%%%%%%%

\section{Constraints on the interior and atmospheric composition of the planet}

\begin{figure*} \includegraphics[width=\textwidth,angle=0]{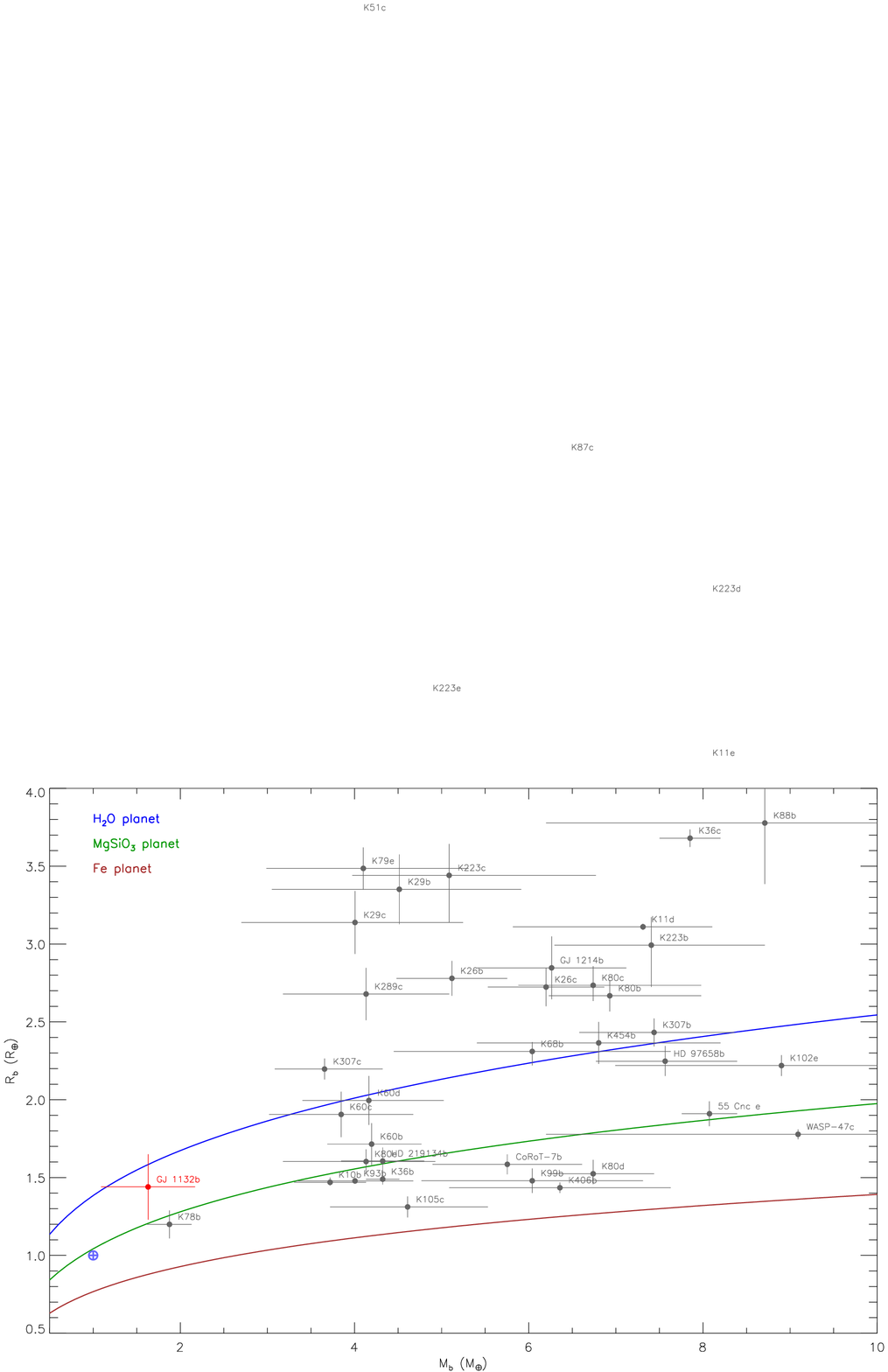}
\caption{\label{fig:madhu:mr} \reff{Mass-radius plot showing the properties of GJ\,1132\,b as well
as values for other planets taken from literature sources and compiled in TEPCat \citep{Me11mn}.
All planets in this mass range are included if their mass is measured to 3$\sigma$ significance
(i.e.\ the upper and lower errorbars on their mass measurements are both less than one third of
the mass value). The names of individual planets are noted, and ``K'' has been used as shorthand
for ``Kepler'' to aid the clarity of the plot. The coloured curves show the mass-radius relation
for planets composed of pure water (blue), enstatite (green) and iron (brown). The position of
the Earth is shown with a $\oplus$ symbol.}} \end{figure*}

\begin{figure*} \includegraphics[width=\textwidth,angle=0]{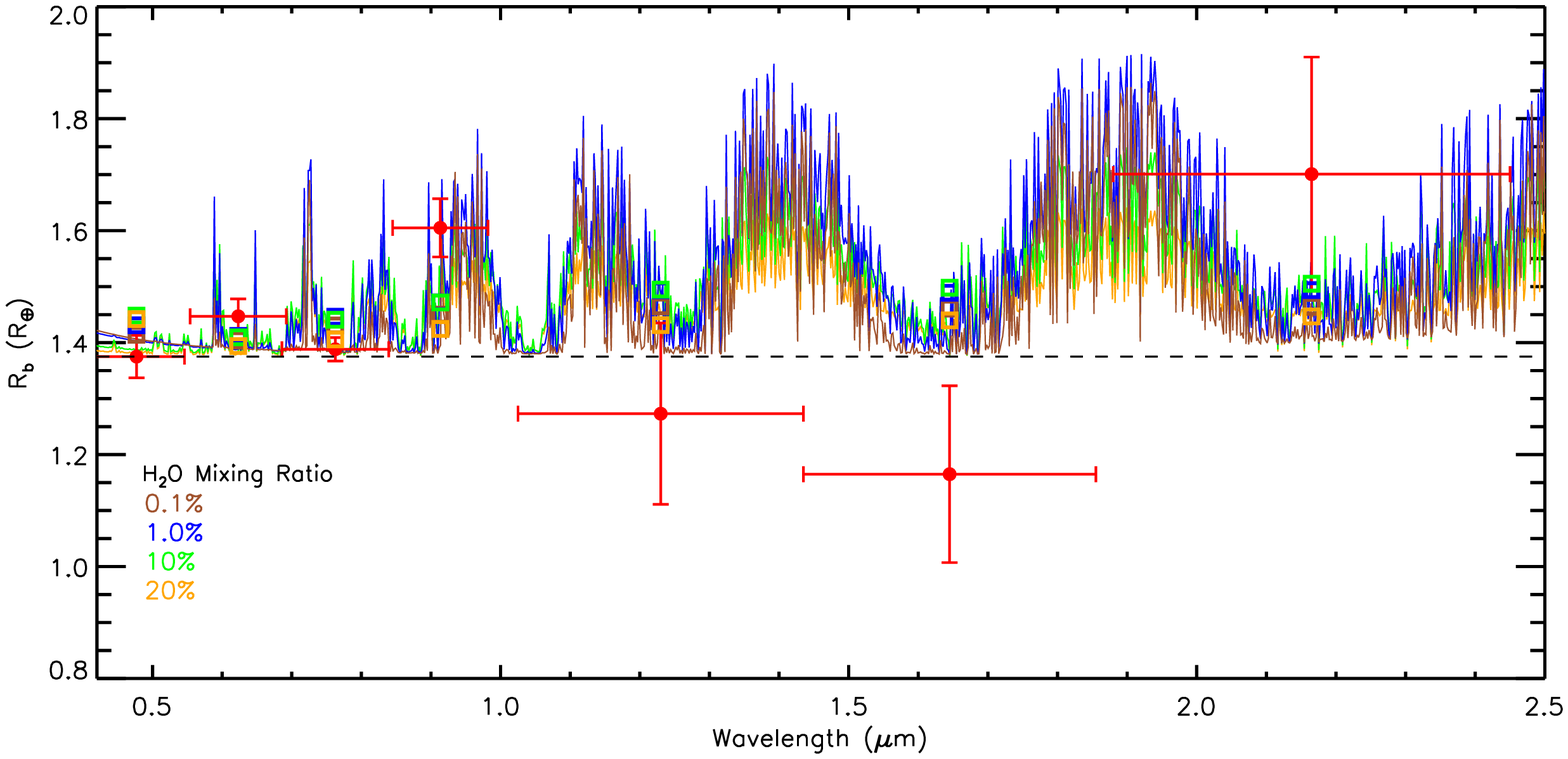}
\caption{\label{fig:madhu:ts} Comparison between the observed transmission spectrum
of GJ\,1132\,b and theoretical spectra for a range of H$_2$O volume mixing ratios
(coloured lines) in a H$_2$-dominated atmosphere. The red points show results from
our GROND observations \reff{and the coloured circles indicate the band-integrated
values of the theoretical spectra. The dashed black line shows the baseline radius
of the planet.}} \end{figure*}

Our multi-band photometric observations allow us to place joint constraints on the interior and atmospheric composition of GJ\,1132\,b. As discussed above our observations include measurements of transit depth, and hence the planetary radius, in seven photometric bands. As shown in Fig.\,\ref{fig:rvary}, six of these seven measurements are consistent with a ``surface radius'' of 1.375\Rearth\ to within the $\sim$2$\sigma$ uncertainties. By ``surface radius'' we mean the smallest radius which is observationally accessible. However, the measurement in the $z$-band at 0.9\,$\mu$m differs from a flat spectrum by over 4$\sigma$, thereby making it highly statistically significant, and suggests a potential contribution from the planetary atmosphere. Additionally, the $z$-band also overlaps with a strong H$_2$O absorption band whereas the remaining six bands probe windows in the H$_2$O opacity. Therefore, it is possible that the latter six bands are all measuring the radius of the planetary ``surface'' while the $z$-band is probing a potentially H$_2$O-rich atmosphere contributing to a higher measured radius. The importance of using radii measured in different bands to represent the interior versus atmosphere has been suggested previously by \citet{MadhusudhanRedfield15ijasb}, an effect which we are likely witnessing in the present case. Therefore, in what follows, we use a combination of interior and atmospheric models to jointly interpret the data.

The observed mass and radius of GJ\,1132\,b allow us to investigate the possible interior composition of the planet using internal structure models. The planet mass as shown in Table\,\ref{tab:model} is $1.63\pm0.54$\Mearth. For the radius, we use the baseline value of $1.375\pm0.16$\Rearth\ discussed above to represent the bulk `grey' radius of the planet. Fig.\,\ref{fig:madhu:mr} shows model mass-radius curves of homogeneous super-Earths of different compositions spanning Fe, silicates, and H$_2$O. The models are described in \citet{Madhusudhan++12apj}. We find that the mass and radius are consistent with two broad compositional regimes. Firstly, an exactly  Earth-like composition, with 33\% iron, 67\% silicates and no volatile layer, is inconsistent with the data within the 1$\sigma$ uncertainties. But, a composition with higher silicate-to-iron fraction, including a pure silicate planet, is ostensibly consistent with the data, albeit marginally. On the other hand, the data are also consistent with a large range of H$_2$O mass fractions between 0\% and 100\% in our models. In principle, consideration of temperature-dependent internal structure models would lead to larger model radii for the same composition \citep{ThomasMadhusudhan16mn} and therefore could lower the upper limit on the water mass fraction. Nevertheless, the mass and radius of GJ\,1132\,b allow for a degenerate set of solutions ranging between a purely silicate bare-rock planet and an ocean planet with a substantial H$_2$O envelope. The degeneracy between the two scenarios could potentially be resolved using spectroscopic observations of the planetary atmosphere, as discussed below.

Considering the transmission spectrum of the planet, we report a tentative inference of H$_2$O in the planetary atmosphere which provides initial signs of a water-rich world. Fig.\,\ref{fig:madhu:ts} shows the data and model transmission spectra of GJ\,1132\,b with different H$_2$O mixing ratios in a H$_2$-rich atmosphere; other compositions are explored in the following section. The data are inconsistent with a flat spectrum by over 4$\sigma$. As discussed above, the key constraint on the atmospheric composition of the planet is governed by the $z$-band measurement which shows a substantially higher transit depth relative to the baseline. We model the atmospheric transmission spectrum of the planet using the exoplanetary atmospheric modelling method of \citet{MadhusudhanSeager09apj} and \citet{Madhusudhan12apj}. Given the limited number of datapoints we did not embark on a full retrieval exercise, but instead systematically investigated a grid of model compositions. In this section, we consider models comprising only gaseous H$_2$ and H$_2$O at different mixing ratios to explore the potential contribution of H$_2$O to the $z$-band measurement. We nominally fixed the temperature structure to be isothermal at 600\,K, representative of the equilibrium temperature of the planet; nominal variations in the temperature don't change our  conclusions significantly. We find that the best model fits to the data, particularly the $z$-band point, are obtained for H$_2$O volume mixing ratios of 1--10\%, as shown in Fig.\,\ref{fig:madhu:ts}. On the other hand, the remaining data which are consistent with a flat spectrum are fit relatively easily for a wide range of models, as the corresponding photometric bands largely probe windows in opacity.

\reff{The amplitude of the observed $z$-band feature is physically plausible for a range of compositions. The difference between the $z$-band radius and the continuum provides an estimate of the thickness of the observable atmosphere ($H_{\rm atm}$) to be $0.22\pm0.06$\Rearth\ or $\sim$$1400\pm400$\,km. Considering an isothermal atmosphere at the equilibrium temperature (644\,K) the atmospheric scale height ($H_{\rm sc}$) for a H$_2$-dominated atmosphere (mean molecular mass $\mu=2$) is $\sim$300\,km. The value of $H_{\rm sc}$ for 10\% H$_2$O ($\mu=3.6$) and 20\% H$_2$O ($\mu=5.2$) is 167\,km and 115\,km, respectively, assuming H$_2$ occupies the remaining fraction. Therefore, the atmospheric height can be explained by $\sim$5 scale heights of an H$_2$-dominated atmosphere given a strong absorber in the $z$-band. Similarly, $H_{\rm atm}$ can also be explained by $\lesssim$10 scale heights of a 10\% H$_2$O atmosphere. Conversely, considering $\sim$8--10 scale heights expected for a saturated spectral feature \citep{MadhusudhanRedfield15ijasb} at the same temperature, the mean molecular mass of the atmosphere is constrained to be $\mu \sim 2.8$--5.5. Such a $\mu$ is possible in a H$_2$-rich atmosphere with $\sim$10--20\% H$_2$O or corresponding fractions of other heavy molecules, again assuming that a strong absorber in the $z$-band is present in the atmosphere. Future observations with HST and JWST in the near-infrared could further constrain the presence and composition of such an absorber.}

The combined interior and atmosphere models fitting the data are consistent with a water-rich envelope in GJ\,1132\,b. And, given that the data are inconsistent with a flat spectrum the atmosphere is unlikely to be of very high mean molecular mass, e.g. 100\% H$_2$O or CO$_2$, or one with thick high altitude clouds. However, since the inference is based primarily on one photometric datapoint (albeit obtained from light curves of six transits), future observations are critical to validate our current finding. We predict that a transmission spectrum of GJ\,1132\,b observed with HST/WFC3 in the 1.1--1.8\,$\mu$m spectral range should be able to detect the water absorption in its atmosphere. On the contrary, if such a HST spectrum does not detect an H$_2$O feature then it might suggest the presence of another molecule in the atmosphere which we have not accounted for in our present models, or the possibility of time-variable events in the planetary atmosphere. It is also necessary that additional observations of the planetary transit in the $z$-band be conducted to bolster the current finding.

\subsection{Transmission spectra calculations using \emph{petitCODE}}

\begin{figure*} \includegraphics[width=\textwidth,angle=0]{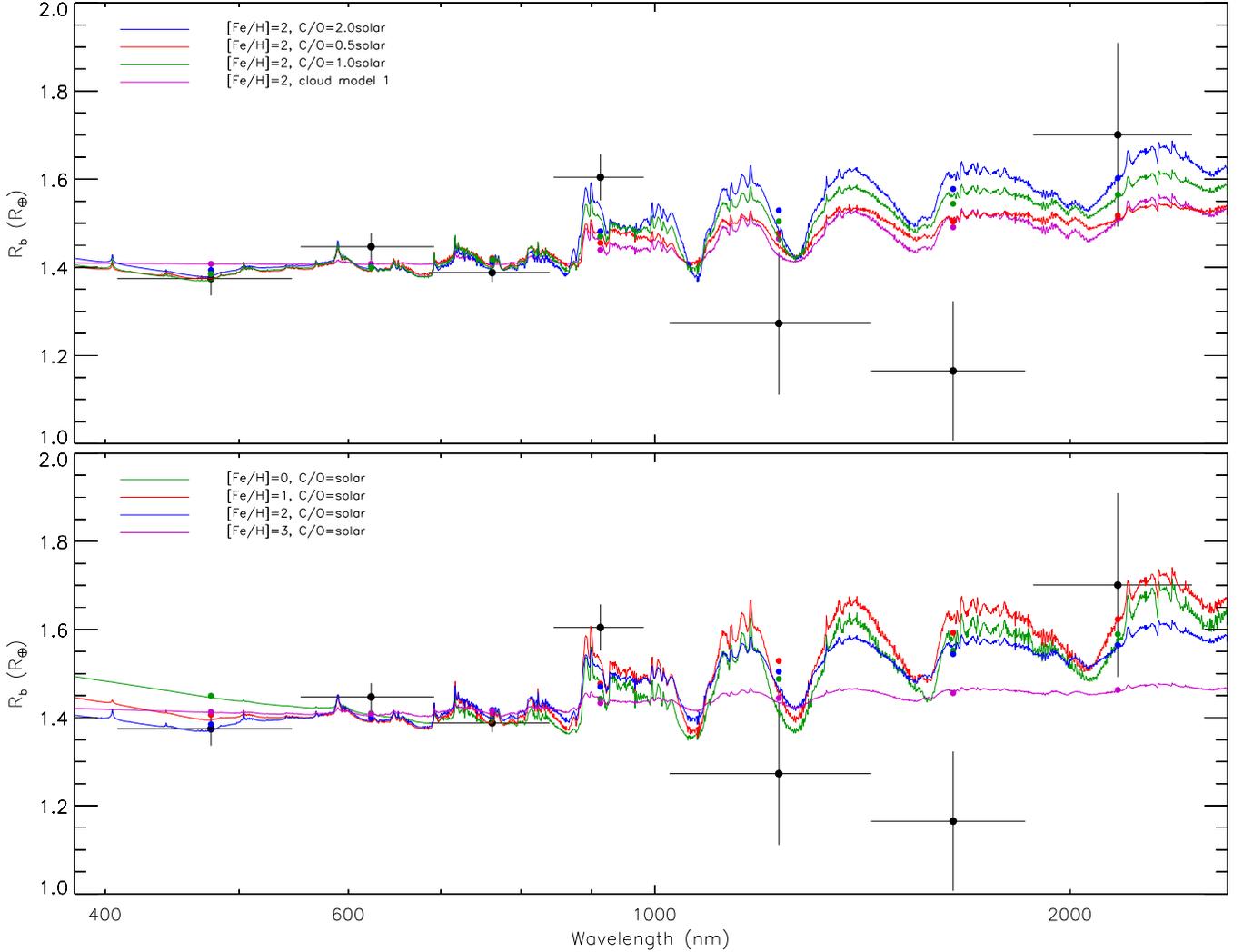}
\caption{\label{fig:mol} \reff{Comparison between the transmission spectrum of GJ\,1132\,b
observed using GROND, and theoretical spectra from \emph{petitCODE}. Black points show the
measured values and coloured points the band-integrated theoretical values of the planetary
radius. The top panel shows the four investigated values of \FeH, and the bottom panel shows
models with $\FeH=2$ and different choices of C/O ratio or treatment of cloud.}} \end{figure*}

Using the newest version of the \emph{petitCODE} \citep{Molliere+15apj,Molliere+17aa}, we performed a second, independent, exploration of the atmospheric properties of the planet. \emph{petitCODE} self-consistently calculates the radiative-convective equilibrium structures of irradiated or self-luminous exoplanet atmospheres. Molecular and atomic line opacities, cloud opacities, and H$_2$--H$_2$ and H$_2$--He collision induced absorption (CIA) are taken into account. The code also includes molecular Rayleigh and cloud particle scattering.

For these calculations we used the planetary parameters as defined in Table\,\ref{tab:model}. We assumed a stellar effective temperature of 3270\,K from BT15. We calculated our standard suite of models as defined in \citet{Molliere+17aa} for this planet, consisting of one fiducial model without clouds, two models at half and twice the solar C/O ratio, respectively, and nine different cloud model approaches as defined in table\,2 of \citet{Molliere+17aa}. For all models (three clear and nine cloudy) we calculated atmospheric structures and spectra at four different scaled-solar chemical compositions, corresponding to atmospheric metallicities of $\FeH=0$, 1, 2 and 3.

We considered Na$_2$S and KCl clouds only, because higher-temperature cloud species can most likely not be mixed up to the higher layers of the atmosphere \citep{Charnay++15apj,Parmentier+16apj}. Cloud models 1 to 4 represent models of varying cloud thickness using the model by \citet{AckermanMarley01apj} and adopting $f_{\rm sed}$ values of 0.01 to 3, where $f_{\rm sed}$ is the ratio of the particle-mass-averaged settling speed and the atmospheric mixing velocity. Cloud models 5 to 9 represent extended clouds with mono-dispersed size distributions. The particle size is fixed at 0.08\,$\mu$m, which leads to Rayleigh scattering of the clouds in the optical. The cloud mass fraction is equal to the mass fraction derived from equilibrium chemistry, but not larger than a maximum value $X_{\rm max}$, which decreases from model 5 to 7. $X_{\rm max}$ can thus be thought of as a simple parametrisation of the settling strength. Cloud models 8 and 9 adopt the same parameter choice as cloud model 6, but additionally include iron clouds (model 8) or a spherical, homogeneous cloud particle shape (model 9). The cloud models therefore cover the parameter space of larger-particle clouds (models 1 to 4), leading to flatter transmission spectra in the optical, to small-particle clouds (models 5 to 9) which lead to Rayleigh scattering. Model 8 will not exhibit Rayleigh scattering if iron can condense within the atmosphere, due to the strong optical absorption of iron.

Each of the model spectra was compared to the observed transmission spectrum and the level of agreement determined. This was done by identifying the lowest $\chi^2$ value between the observed transmission spectra and passband-integrated model values from \emph{petitCODE}, whilst allowing for an overall shift in radius because the spectra were calculated assuming a planetary base radius of 1.43\Rearth\ at 10\,bar pressure. Faced with a multitude of choices over which data to consider, we defaulted to using the GROND $griz$ points with one limb darkening coefficient fitted and the GROND $JHK$ points with fixed limb darkening coefficients. This is because there is a disagreement between the planet radius in the GROND and PISCO $g$ bands, and the former was obtained from nine transits versus the single transit for the latter. We prefer results obtained with fitted limb darkening coefficients because of limitations in the theoretical understanding of the atmospheres of stars as cool as GJ\,1132\,A. Alternative choices of datapoints will be discussed below. Selected models are shown in Fig.\,\ref{fig:mol}.

\reff{We found that the best fit to the GROND $grizJHK$ transmission spectrum is obtained for the \emph{petitCODE} model with $\FeH=2$ and a twice-solar or solar C/O ratio ($\chi^2=18.6$ and 19.1, respectively, for seven datapoints) although acceptable agreement also occurs for the models with $\FeH=1$ and $\FeH=3$ ($\chi^2$ from 19.9 to 20.7 depending on C/O ratio). The models with $\FeH=3$ have muted spectral features, because the high mean molecular mass leads to a low atmospheric scale height. They match the $H$-band radius better, at the expense of a poorer agreement with the $z$-band radius. A featureless spectrum (i.e.\ a straight line) is a worse fit with $\chi^2=22.1$. The $\chi^2$ values are all quite large, and reflect the difficulty of finding a good agreement between the observed and theoretical spectra. Including cloud in the models serves to increase the $\chi^2$ because the flattened spectral features match the observed transmission spectrum less well.}

\reff{If we restrict the analysis to the GROND $griz$ datapoints, as these are the most reliable measurements, the models with $\FeH=1$ and $\FeH=2$ are the best match. The greatest agreements are for $\FeH=2$ and a twice-solar C/O ratio ($\chi^2=8.9$ for four datapoints), followed by $\FeH=1$ and a solar ($\chi^2=10.0$) or twice-solar ($\chi^2=10.5$) C/O ratio. The straight-line fit is clearly disfavoured with $\chi^2=16.9$. The slight preference for the twice-solar C/O ratio is because a higher C/O ratio enhances the amount of CH$_4$, an effect which outweighs the loss of H$_2$O and so leads to a larger radius in the $z$ band. Even these models underpredict the detected variation of planet radius with wavelength, which also means a cloudy or featureless spectrum is strongly disfavoured. We therefore conclude that our results are best explained by a clear atmosphere with $\FeH=2$ and strong $z$-band opacity contributed by CH$_4$ or H$_2$O.}

\reff{Turning now to the transmission spectrum from the GROND data with all limb darkening coefficients fixed, we find that the best models have $\FeH=2$, with $\chi^2=19.4$ (seven dataponts) for a solar C/O ratio, 19.5 for twice-solar C/O and 19.8 for half-solar C/O. $\FeH=3$ with twice-solar C/O is nearly as good ($\chi^2=20.2$), as is $\FeH=1$ with half-solar C/O ($\chi^2=20.8$). A straight-line fit has $\chi^2=22.1$. The quality of our data and limitations in the treatment of limb darkening means that we are able to infer that the atmosphere of GJ\,1132\,b is best-represented by models with $\FeH=2$, but the C/O ratio is not usefully constrained.}

%%%%%%%%%%%%%%%%%%%%%%%%%%%%%%%%%%%%%%%%%%%%%%%%%%%%%%%%%%%%%%%%%%%%%%%%%%%%%%%%%%%%%%%%%%%%%%%%%%%%%%%%%%%%%%%%%%%%%%%%%%%%%%%%%%%%%%%%%%%%%%%%%%%%%

\section{Summary and conclusions}

GJ\,1132 is a \reff{benchmark} nearby system containing a low-mass planet transiting a late-M dwarf. We have presented extensive photometry of the system comprising light curves of nine transits observed simultaneously in the $griz$ optical and $JHK$ near-IR passbands. We have analysed these and literature data to determine the physical properties of the system. We find that the planet is larger than previously thought, $1.43\pm0.16$\Rearth\ versus $1.16\pm0.11$\Rearth, from a methodological approach which relies more on observations of the system and less on empirical calibrations of the properties of low-mass stars. The planet's measured mass and radius are consistent within 1$\sigma$ with theoretical predictions for a planet composed of silicates or water; a 100\% iron composition gives a radius too small by $\sim$2$\sigma$.

Our repeat observations allowed us to check for variability in the measured planet radius, and two of the transits do indeed yield radii which are modestly discrepant with measurements from other transits. This could indicate excess scatter among the results, starspots on the stellar surface, unidentified systematic effects in our data, or the presence of \reff{variability in the planet's atmosphere \citep[e.g.][]{Armstrong+16natas}.}

We have constructed an optical-\reff{infrared} transmission spectrum of GJ\,1132\,b by modelling all light curves with a consistent geometry. We find an increased planet radius in the $z$ band, to a significance level of 4$\sigma$, indicative of atmospheric opacity due to water, \reff{methane or another unidentified source}. Detailed investigation of the resulting errorbars was enabled by the observation of nine transits. We find that our results are robust against the rejection of individual transits or the inclusion of contaminating light from a nearby star. The treatment of limb darkening is more concerning, as it affects the results in the $r$ and $i$ bands at the level of 2.7$\sigma$ and 3.5$\sigma$, respectively. We urge fellow researchers to consider this issue in similar analyses, especially for very cool stars where theoretical limb darkening coefficients are less reliable.

The transmission spectrum was modelled using the atmospheric models of \citet{MadhusudhanSeager09apj}, with the finding that H$_2$O likely causes the enlarged $z$-band radius of the planet. The best fits to the observations are found for H$_2$O volume mixing ratios of 1--10\%, implying a water-rich atmospheric composition which would cause observable spectral features in a 1.1--1.8\,$\mu$m transmission spectrum obtained using HST/WFC3. From simulations of the atmosphere of GJ\,1132\,b, \citet{Schaefer+16apj} found that the presence of H$_2$O implied either an H$_2$ envelope or low UV flux from the host star early in the lifetime of the system, and the ongoing presence of a magma ocean on the planet's surface.

We also calculated theoretical spectra using the \emph{petitCODE} \citep{Molliere+15apj}, which yield similar results except for the finding that the large $z$-band radius is explicable by an enhanced abundance of CH$_4$. A high metallicity of $\FeH=2$ is preferred, depending on the datapoints considered, which is in line with the mass--metallicity correlation seen for more massive planets \citep{Kreidberg+14apj,Mordasini+16apj}. A straight line is a much poorer fit to the transmission spectrum, confirming that we have detected the atmosphere of a 1.6\Mearth\ planet.

We advocate extensive further observations to refine and extend our understanding of the GJ\,1132 system. High-precision optical light curves from large telescopes would be able to confirm or disprove the larger radius of the planet in the $z$-band, and shed light on the discrepancy seen in the $g$-band. Intermediate-band photometry at 900\,nm or bluer than 500\,nm would enable finer distinctions to be made between competing model spectra and a clearer understanding of the chemical composition of the planetary atmosphere. The planet's mean density measurement is also hindered by the weak detection of the velocity motion of the host star, an issue which could be ameliorated with further radial velocity measurements using large telescopes. Finally, infrared transit photometry and spectroscopy should allow the detection of a range of molecules via the absorption features they imprint on the spectrum of the planet's atmosphere as backlit by its host star.

Our results show that a 1.6\Mearth\ planet with an equilibrium temperature of 650\,K is capable of retaining an extensive atmosphere. The atmosphere contains multiple molecular species and has likely persisted for many Gyr since the formation of the system.

%%%%%%%%%%%%%%%%%%%%%%%%%%%%%%%%%%%%%%%%%%%%%%%%%%%%%%%%%%%%%%%%%%%%%%%%%%%%%%%%%%%%%%%%%%%%%%%%%%%%%%%%%%%%%%%%%%%%%%%%%%%%%%%%%%%%%%%%%%%%%%%%%%%%%

\section*{Acknowledgements}

JS acknowledges financial support from the Leverhulme Trust in the form of a Philip Leverhulme Prize. We thank BT15 for making their photometric data available. The following internet-based resources were used in research for this paper: the ESO Digitized Sky Survey; the NASA Astrophysics Data System; the SIMBAD database and VizieR catalogue access tool operated at CDS, Strasbourg, France; and the ar$\chi$iv scientific paper preprint service operated by Cornell University.

%%%%%%%%%%%%%%%%%%%%%%%%%%%%%%%%%%%%%%%%%%%%%%%%%%%%%%%%%%%%%%%%%%%%%%%%%%%%%%%%%%%%%%%%%%%%%%%%%%%%%%%%%%%%%%%%%%%%%%%%%%%%%%%%%%%%%%%%%%%%%%%%%%%%%

\bibliographystyle{mn_new}
% \bibliography{aamnem99,jkt}

\begin{thebibliography}{81}
\expandafter\ifx\csname natexlab\endcsname\relax\def\natexlab#1{#1}\fi

\bibitem[{{Ackerman} \& {Marley}(2001)}]{AckermanMarley01apj}
{Ackerman}, A.~S., {Marley}, M.~S., 2001, ApJ, 556, 872

\bibitem[{{Alonso} et~al.(2008){Alonso}, {Barbieri}, {Rabus}, {Deeg},
  {Belmonte}, \& {Almenara}}]{Alonso+08aa}
{Alonso}, R., {Barbieri}, M., {Rabus}, M., {Deeg}, H.~J., {Belmonte}, J.~A.,
  {Almenara}, J.~M., 2008, A\&A, 487, L5

\bibitem[{{Anglada-Escud{\'e}} et~al.(2013){Anglada-Escud{\'e}}, {Rojas-Ayala},
  {Boss}, {Weinberger}, \& {Lloyd}}]{Anglada+13aa}
{Anglada-Escud{\'e}}, G., {Rojas-Ayala}, B., {Boss}, A.~P., {Weinberger},
  A.~J., {Lloyd}, J.~P., 2013, A\&A, 551, A48

\bibitem[{{Armstrong} et~al.(2016){Armstrong}, {de Mooij}, {Barstow}, {Osborn},
  {Blake}, \& {Saniee}}]{Armstrong+16natas}
{Armstrong}, D.~J., {de Mooij}, E., {Barstow}, J., {Osborn}, H.~P., {Blake},
  J., {Saniee}, N.~F., 2016, Nature Astronomy, 1, 0004

\bibitem[{{Awiphan} et~al.(2016)}]{Awiphan+16mn}
{Awiphan}, S., et~al., 2016, MNRAS, 463, 2574

\bibitem[{{Ballerini} et~al.(2012){Ballerini}, {Micela}, {Lanza}, \&
  {Pagano}}]{Ballerini+12aa}
{Ballerini}, P., {Micela}, G., {Lanza}, A.~F., {Pagano}, I., 2012, A\&A, 539,
  A140

\bibitem[{{Bean} et~al.(2010){Bean}, {Miller-Ricci Kempton}, \&
  {Homeier}}]{Bean+10nat}
{Bean}, J.~L., {Miller-Ricci Kempton}, E., {Homeier}, D., 2010, Nature, 468,
  669

\bibitem[{{Bean} et~al.(2011)}]{Bean+11apj}
{Bean}, J.~L., et~al., 2011, ApJ, 743, 92

\bibitem[{{Berta-Thompson} et~al.(2015)}]{Berta+15nat}
{Berta-Thompson}, Z.~K., et~al., 2015, Nature, 527, 204

\bibitem[{{Biddle} et~al.(2014)}]{Biddle+14mn}
{Biddle}, L.~I., et~al., 2014, MNRAS, 443, 1810

\bibitem[{{Bonfils} et~al.(2012)}]{Bonfils+12aa}
{Bonfils}, X., et~al., 2012, A\&A, 546, A27

\bibitem[{{Cassan} et~al.(2012)}]{Cassan+12nat}
{Cassan}, A., et~al., 2012, Nature, 481, 167

\bibitem[{{Charbonneau} et~al.(2009)}]{Charbonneau+09nat}
{Charbonneau}, D., et~al., 2009, Nature, 462, 891

\bibitem[{{Charnay} et~al.(2015){Charnay}, {Meadows}, \&
  {Leconte}}]{Charnay++15apj}
{Charnay}, B., {Meadows}, V., {Leconte}, J., 2015, ApJ, 813, 15

\bibitem[{{Chen} et~al.(2014){Chen}, {van Boekel}, {Madhusudhan}, {Wang},
  {Nikolov}, {Seemann}, \& {Henning}}]{Chen+14aa}
{Chen}, G., {van Boekel}, R., {Madhusudhan}, N., {Wang}, H., {Nikolov}, N.,
  {Seemann}, U., {Henning}, T., 2014, A\&A, 564, A6

\bibitem[{{Claret}(2004)}]{Claret04aa2}
{Claret}, A., 2004, A\&A, 428, 1001

\bibitem[{{Claret} et~al.(2012){Claret}, {Hauschildt}, \&
  {Witte}}]{Claret++12aa}
{Claret}, A., {Hauschildt}, P.~H., {Witte}, S., 2012, A\&a, 546, A14

\bibitem[{{Cloutier} et~al.(2017){Cloutier}, {Doyon}, {Menou}, {Delfosse},
  {Dumusque}, \& {Artigau}}]{Cloutier+17aj}
{Cloutier}, R., {Doyon}, R., {Menou}, K., {Delfosse}, X., {Dumusque}, X.,
  {Artigau}, {\'E}., 2017, AJ, 153, 9

\bibitem[{{Cunha} et~al.(2015){Cunha}, {Correia}, \& {Laskar}}]{Cunha++14ijasb}
{Cunha}, D., {Correia}, A.~C.~M., {Laskar}, J., 2015, International Journal of
  Astrobiology, 14, 233

\bibitem[{{de Mooij} et~al.(2012)}]{Demooij+12aa}
{de Mooij}, E.~J.~W., et~al., 2012, A\&A, 538, A46

\bibitem[{{Delfosse} et~al.(2000){Delfosse}, {Forveille}, {S{\'e}gransan},
  {Beuzit}, {Udry}, {Perrier}, \& {Mayor}}]{Delfosse+00aa}
{Delfosse}, X., {Forveille}, T., {S{\'e}gransan}, D., {Beuzit}, J.-L., {Udry},
  S., {Perrier}, C., {Mayor}, M., 2000, A\&A, 364, 217

\bibitem[{{Demory} et~al.(2016)}]{Demory+16nat}
{Demory}, B.-O., et~al., 2016, Nature, 532, 207

\bibitem[{{Dittmann} et~al.(2016){Dittmann}, {Irwin}, {Charbonneau},
  {Berta-Thompson}, \& {Newton}}]{Dittmann+16}
{Dittmann}, J.~A., {Irwin}, J.~M., {Charbonneau}, D., {Berta-Thompson}, Z.~K.,
  {Newton}, E.~R., 2016, ApJ, submitted, {\tt arXiv:1611.09848}

\bibitem[{{Dragomir} et~al.(2015){Dragomir}, {Benneke}, {Pearson},
  {Crossfield}, {Eastman}, {Barman}, \& {Biddle}}]{Dragomir+15apj}
{Dragomir}, D., {Benneke}, B., {Pearson}, K.~A., {Crossfield}, I.~J.~M.,
  {Eastman}, J., {Barman}, T., {Biddle}, L.~I., 2015, ApJ, 814, 102

\bibitem[{{Dragomir} et~al.(2013)}]{Dragomir+13apj}
{Dragomir}, D., et~al., 2013, ApJ, 772, L2

\bibitem[{{Dressing} \& {Charbonneau}(2015)}]{DressingCharbonneau15apj}
{Dressing}, C.~D., {Charbonneau}, D., 2015, ApJ, 807, 45

\bibitem[{{Eastman} et~al.(2010){Eastman}, {Siverd}, \&
  {Gaudi}}]{Eastman++10pasp}
{Eastman}, J., {Siverd}, R., {Gaudi}, B.~S., 2010, PASP, 122, 935

\bibitem[{{Fukugita} et~al.(1996){Fukugita}, {Ichikawa}, {Gunn}, {Doi},
  {Shimasaku}, \& {Schneider}}]{Fukugita+96aj}
{Fukugita}, M., {Ichikawa}, T., {Gunn}, J.~E., {Doi}, M., {Shimasaku}, K.,
  {Schneider}, D.~P., 1996, AJ, 111, 1748

\bibitem[{{Gaidos} et~al.(2016){Gaidos}, {Mann}, {Kraus}, \&
  {Ireland}}]{Gaidos+16mn}
{Gaidos}, E., {Mann}, A.~W., {Kraus}, A.~L., {Ireland}, M., 2016, MNRAS, 457,
  2877

\bibitem[{{Gillon} et~al.(2007)}]{Gillon+07aa2}
{Gillon}, M., et~al., 2007, A\&A, 472, L13

\bibitem[{{Gillon} et~al.(2016)}]{Gillon+16nat}
{Gillon}, M., et~al., 2016, Nature, 533, 221

\bibitem[{{Greiner} et~al.(2008)}]{Greiner+08pasp}
{Greiner}, J., et~al., 2008, PASP, 120, 405

\bibitem[{{Hartman} et~al.(2015)}]{Hartman+15aj}
{Hartman}, J.~D., et~al., 2015, AJ, 149, 166

\bibitem[{{Hilditch}(2001)}]{Hilditch01book}
{Hilditch}, R.~W., 2001, {An Introduction to Close Binary Stars}, Cambridge
  University Press, Cambridge, UK

\bibitem[{{Jackson} \& {Jeffries}(2012)}]{JacksonJeffries12mn}
{Jackson}, R.~J., {Jeffries}, R.~D., 2012, MNRAS, 423, 2966

\bibitem[{{Jackson} \& {Jeffries}(2013)}]{JacksonJeffries13mn}
{Jackson}, R.~J., {Jeffries}, R.~D., 2013, MNRAS, 431, 1883

\bibitem[{{Jao} et~al.(2005){Jao}, {Henry}, {Subasavage}, {Brown}, {Ianna},
  {Bartlett}, {Costa}, \& {M{\'e}ndez}}]{Jao+05aj}
{Jao}, W.-C., {Henry}, T.~J., {Subasavage}, J.~P., {Brown}, M.~A., {Ianna},
  P.~A., {Bartlett}, J.~L., {Costa}, E., {M{\'e}ndez}, R.~A., 2005, AJ, 129,
  1954

\bibitem[{{Knutson} et~al.(2014{\natexlab{a}}){Knutson}, {Benneke}, {Deming},
  \& {Homeier}}]{Knutson+14nat}
{Knutson}, H.~A., {Benneke}, B., {Deming}, D., {Homeier}, D.,
  2014{\natexlab{a}}, Nature, 505, 66

\bibitem[{{Knutson} et~al.(2014{\natexlab{b}})}]{Knutson+14apj2}
{Knutson}, H.~A., et~al., 2014{\natexlab{b}}, ApJ, 794, 155

\bibitem[{{Kopparapu} et~al.(2013)}]{Kopparapu+13apj}
{Kopparapu}, R.~K., et~al., 2013, ApJ, 765, 131

\bibitem[{{Kreidberg} et~al.(2014{\natexlab{a}})}]{Kreidberg+14apj}
{Kreidberg}, L., et~al., 2014{\natexlab{a}}, ApJ, 793, L27

\bibitem[{{Kreidberg} et~al.(2014{\natexlab{b}})}]{Kreidberg+14nat}
{Kreidberg}, L., et~al., 2014{\natexlab{b}}, Nature, 505, 69

\bibitem[{{Lammer} et~al.(2003){Lammer}, {Selsis}, {Ribas}, {Guinan}, {Bauer},
  \& {Weiss}}]{Lammer+03apj}
{Lammer}, H., {Selsis}, F., {Ribas}, I., {Guinan}, E.~F., {Bauer}, S.~J.,
  {Weiss}, W.~W., 2003, ApJ, 598, L121

\bibitem[{{Lanotte} et~al.(2014)}]{Lanotte+14aa}
{Lanotte}, A.~A., et~al., 2014, A\&A, 572, A73

\bibitem[{{Madhusudhan}(2012)}]{Madhusudhan12apj}
{Madhusudhan}, N., 2012, ApJ, 758, 36

\bibitem[{{Madhusudhan} \& {Redfield}(2015)}]{MadhusudhanRedfield15ijasb}
{Madhusudhan}, N., {Redfield}, S., 2015, International Journal of Astrobiology,
  14, 177

\bibitem[{{Madhusudhan} \& {Seager}(2009)}]{MadhusudhanSeager09apj}
{Madhusudhan}, N., {Seager}, S., 2009, ApJ, 707, 24

\bibitem[{{Madhusudhan} et~al.(2012){Madhusudhan}, {Lee}, \&
  {Mousis}}]{Madhusudhan++12apj}
{Madhusudhan}, N., {Lee}, K.~K.~M., {Mousis}, O., 2012, ApJ, 759, L40

\bibitem[{{Mancini} et~al.(2014)}]{Mancini+14aa}
{Mancini}, L., et~al., 2014, A\&A, 562, A126

\bibitem[{{Molli{\`e}re} et~al.(2015){Molli{\`e}re}, {van Boekel}, {Dullemond},
  {Henning}, \& {Mordasini}}]{Molliere+15apj}
{Molli{\`e}re}, P., {van Boekel}, R., {Dullemond}, C., {Henning}, T.,
  {Mordasini}, C., 2015, ApJ, 813, 47

\bibitem[{{Molli{\`e}re} et~al.(2016){Molli{\`e}re}, {van Boekel}, {Bouwman},
  {Henning}, {Lagage}, \& {Min}}]{Molliere+17aa}
{Molli{\`e}re}, P., {van Boekel}, R., {Bouwman}, J., {Henning}, T., {Lagage},
  P.-O., {Min}, M., 2016, A\&A, in press, {\tt arXiv:1611.08608}

\bibitem[{{Monet} et~al.(2003)}]{Monet+03aj}
{Monet}, D.~G., et~al., 2003, AJ, 125, 984

\bibitem[{{Mordasini} et~al.(2016){Mordasini}, {van Boekel}, {Molli{\`e}re},
  {Henning}, \& {Benneke}}]{Mordasini+16apj}
{Mordasini}, C., {van Boekel}, R., {Molli{\`e}re}, P., {Henning}, T.,
  {Benneke}, B., 2016, ApJ, 832, 41

\bibitem[{{Morton} \& {Swift}(2014)}]{MortonSwift14apj}
{Morton}, T.~D., {Swift}, J., 2014, ApJ, 791, 10

\bibitem[{{Muirhead} et~al.(2015)}]{Muirhead+15apj}
{Muirhead}, P.~S., et~al., 2015, ApJ, 801, 18

\bibitem[{{Nascimbeni} et~al.(2013){Nascimbeni}, {Piotto}, {Pagano},
  {Scandariato}, {Sani}, \& {Fumana}}]{Nascimbeni+13aa2}
{Nascimbeni}, V., {Piotto}, G., {Pagano}, I., {Scandariato}, G., {Sani}, E.,
  {Fumana}, M., 2013, A\&A, 559, A32

\bibitem[{{Oshagh} et~al.(2013){Oshagh}, {Santos}, {Boisse}, {Bou{\'e}},
  {Montalto}, {Dumusque}, \& {Haghighipour}}]{Oshagh+13aa}
{Oshagh}, M., {Santos}, N.~C., {Boisse}, I., {Bou{\'e}}, G., {Montalto}, M.,
  {Dumusque}, X., {Haghighipour}, N., 2013, A\&A, 556, A19

\bibitem[{{Oshagh} et~al.(2014){Oshagh}, {Santos}, {Ehrenreich},
  {Haghighipour}, {Figueira}, {Santerne}, \& {Montalto}}]{Oshagh+14aa}
{Oshagh}, M., {Santos}, N.~C., {Ehrenreich}, D., {Haghighipour}, N.,
  {Figueira}, P., {Santerne}, A., {Montalto}, M., 2014, A\&A, in press, {\tt
  {arXiv:1407.2066}}

\bibitem[{{Parmentier} et~al.(2016){Parmentier}, {Fortney}, {Showman},
  {Morley}, \& {Marley}}]{Parmentier+16apj}
{Parmentier}, V., {Fortney}, J.~J., {Showman}, A.~P., {Morley}, C., {Marley},
  M.~S., 2016, ApJ, 828, 22

\bibitem[{{Rogers}(2015)}]{Rogers15apj}
{Rogers}, L.~A., 2015, ApJ, 801, 41

\bibitem[{{Schaefer} et~al.(2016){Schaefer}, {Wordsworth}, {Berta-Thompson}, \&
  {Sasselov}}]{Schaefer+16apj}
{Schaefer}, L., {Wordsworth}, R., {Berta-Thompson}, Z., {Sasselov}, D., 2016,
  ApJ, 829, 63

\bibitem[{{Seager} \& {Mall{\'e}n-Ornelas}(2003)}]{SeagerMallen03apj}
{Seager}, S., {Mall{\'e}n-Ornelas}, G., 2003, ApJ, 585, 1038

\bibitem[{{Shields} et~al.(2016){Shields}, {Ballard}, \&
  {Johnson}}]{Shields+16}
{Shields}, A.~L., {Ballard}, S., {Johnson}, J.~A., 2016, {\tt arXiv:1610.05765}

\bibitem[{{Southworth}(2008)}]{Me08mn}
{Southworth}, J., 2008, MNRAS, 386, 1644

\bibitem[{{Southworth}(2010)}]{Me10mn}
{Southworth}, J., 2010, MNRAS, 408, 1689

\bibitem[{{Southworth}(2011)}]{Me11mn}
{Southworth}, J., 2011, MNRAS, 417, 2166

\bibitem[{{Southworth}(2013)}]{Me13aa}
{Southworth}, J., 2013, A\&A, 557, A119

\bibitem[{{Southworth}(2017)}]{Me17pasp}
{Southworth}, J., 2017, PASP, 129, 024401

\bibitem[{{Southworth} \& {Evans}(2016)}]{MeEvans16mn}
{Southworth}, J., {Evans}, D.~F., 2016, MNRAS, 463, 37

\bibitem[{{Southworth} et~al.(2004){Southworth}, {Maxted}, \&
  {Smalley}}]{Me++04mn2}
{Southworth}, J., {Maxted}, P.~F.~L., {Smalley}, B., 2004, MNRAS, 351, 1277

\bibitem[{{Southworth} et~al.(2012){Southworth}, {Mancini}, {Maxted}, {Bruni},
  {Tregloan-Reed}, {Barbieri}, {Ruocco}, \& {Wheatley}}]{Me+12mn2}
{Southworth}, J., {Mancini}, L., {Maxted}, P.~F.~L., {Bruni}, I.,
  {Tregloan-Reed}, J., {Barbieri}, M., {Ruocco}, N., {Wheatley}, P.~J., 2012,
  MNRAS, 422, 3099

\bibitem[{{Southworth} et~al.(2009)}]{Me+09mn}
{Southworth}, J., et~al., 2009, MNRAS, 396, 1023

\bibitem[{{Southworth} et~al.(2014)}]{Me+14mn}
{Southworth}, J., et~al., 2014, MNRAS, 444, 776

\bibitem[{{Sozzetti} et~al.(2007){Sozzetti}, {Torres}, {Charbonneau}, {Latham},
  {Holman}, {Winn}, {Laird}, \& {O'Donovan}}]{Sozzetti+07apj}
{Sozzetti}, A., {Torres}, G., {Charbonneau}, D., {Latham}, D.~W., {Holman},
  M.~J., {Winn}, J.~N., {Laird}, J.~B., {O'Donovan}, F.~T., 2007, ApJ, 664,
  1190

\bibitem[{{Stetson}(1987)}]{Stetson87pasp}
{Stetson}, P.~B., 1987, PASP, 99, 191

\bibitem[{{Thomas} \& {Madhusudhan}(2016)}]{ThomasMadhusudhan16mn}
{Thomas}, S.~W., {Madhusudhan}, N., 2016, MNRAS, 458, 1330

\bibitem[{{Tregloan-Reed} et~al.(2013){Tregloan-Reed}, {Southworth}, \&
  {Tappert}}]{Tregloan++13mn}
{Tregloan-Reed}, J., {Southworth}, J., {Tappert}, C., 2013, MNRAS, 428, 3671

\bibitem[{{Tsiaras} et~al.(2016)}]{Tsiaras+16apj}
{Tsiaras}, A., et~al., 2016, ApJ, 820, 99

\bibitem[{{Van Grootel} et~al.(2014)}]{Vangrootel+14apj}
{Van Grootel}, V., et~al., 2014, ApJ, 786, 2

\bibitem[{{Weiss} \& {Marcy}(2014)}]{WeissMarcy14apjl}
{Weiss}, L.~M., {Marcy}, G.~W., 2014, ApJ, 783, L6

\bibitem[{{Winn} et~al.(2011)}]{Winn+11apj}
{Winn}, J.~N., et~al., 2011, ApJ, 737, L18

\end{thebibliography}

%%%%%%%%%%%%%%%%%%%%%%%%%%%%%%%%%%%%%%%%%%%%%%%%%%%%%%%%%%%%%%%%%%%%%%%%%%%%%%%%%%%%%%%%%%%%%%%%%%%%%%%%%%%%%%%%%%%%%%%%%%%%%%%%%%%%%%%%%%%%%%%%%%%%%
\end{document}